\documentclass[prl,aps,amsfonts,preprint,floats,tightenlines,floatfix]{revtex4}

\usepackage{epsfig}

\begin{document}
\date{\today}
\title{
\hspace*{\fill}{\small\sf UNITU--THEP--06/2002}\\
\hspace*{\fill}{\small\sf http://xxx.lanl.gov/abs/hep-ph/0202195}\\ ~\\ ~\\
The Elusiveness of Infrared Critical Exponents \\
in Landau Gauge Yang--Mills Theories}

\author{
C.~S.~Fischer\footnote{E-Mail: chfi@axion01.tphys.physik.uni-tuebingen.de}, 
R.~Alkofer\footnote{E-Mail: reinhard.alkofer@uni-tuebingen.de}, 
and H.~Reinhardt}
\affiliation{Institute for Theoretical Physics, University of T\"ubingen \\
          Auf der Morgenstelle 14, D-72076 T\"ubingen, Germany}

\begin{abstract}
We solve a truncated system of coupled Dyson--Schwinger equations for the gluon
and ghost propagators in SU($N_c$) Yang--Mills theories in Faddeev--Popov
quantization on a four-torus. This compact space-time manifold provides an
efficient mean to solve the gluon and ghost Dyson--Schwinger equations without
any angular approximations. We verify that analytically two power-like
solutions in the very far infrared seem possible. However, only one of
these solutions can be matched to a numerical solution for non-vanishing
momenta. For a bare ghost-gluon vertex this implies that the gluon propagator
is only weakly infrared vanishing, $D_{gl}(k^2) \propto (k^2)^{2\kappa -1}$,
$\kappa \approx 0.595$, and the ghost propagator is infrared singular,
$D_{gh}(k^2) \propto (k^2)^{-\kappa -1}$. For non-vanishing momenta our
solutions are in agreement with the results of recent SU(2) Monte-Carlo lattice
calculations. The running coupling possesses an infrared fixed point. We obtain
$\alpha(0) = 8.92/N_c$  for all gauge groups SU($N_c$). Above one GeV the
running coupling rapidly approaches its perturbative form.
~
  \\
  Gluon propagator; Dyson--Schwinger equations; Infrared behavior.
\end{abstract}
\maketitle

\newpage

\section{I. Introduction}

It is generally accepted that the theory of strong interactions, QCD, should
describe the observed phenomenon of confinement: colored objects like quarks and
gluons occur only in hadrons. A possible route for gaining more understanding
of this phenomenon is the study of the infrared behavior of QCD Green's
functions, for a recent review see \cite{Alkofer:2001wg}. In addition to shed
light on the fundamental properties of QCD the knowledge of these Green's
functions provides the basis for a successful description of hadronic physics
\cite{Alkofer:2001wg,Roberts:2000aa}. 
Based on the idea of infrared slavery older works on this subject assumed a
strongly infrared singular gluon propagator. Recent studies based either
on Dyson--Schwinger equations
\cite{vonSmekal:1997is,Atkinson:1998tu,Watson:2001yv,Zwanziger:2001kw,Lerche:2001}
or lattice calculations
\cite{Mandula:1999nj,Bonnet:2000kw,Bonnet:2001uh,Langfeld:2001cz}
in Landau gauge indicate quite the opposite: an infrared finite or even 
infrared vanishing gluon propagator.
These two techniques are complementary in the following sense:
On the one hand, Monte-Carlo lattice calculations include all non-perturbative 
physics of  Yang--Mills theories but cannot make definite statements about 
the very far infrared due to the finite lattice volume. On the other hand, 
Dyson--Schwinger equations allow to extract the leading infrared behavior
analytically and the general non-perturbative behavior with moderate numerical
effort but these equations, consisting of an infinite tower of coupled 
non-linear integral equations, have to be truncated in order to be manageable.
As we will also see in the course of this article, the propagators  
of SU(2) and SU(3) Landau gauge Yang--Mills theory in Faddeev--Popov 
quantization coincide for these two different approaches reasonably well.
Thus we are confident that our results for the qualitative features of
these propagators are trustable. 

Especially, these recent results on the Landau gauge propagators imply  that
the Kugo--Ojima confinement criterion
\cite{Kugo:1979gm,Nakanishi:1990qm,Kugo:1995km} is fulfilled
(for a short summary on this topic see {\it e.g.\/} ref.\ 
\cite{vonSmekal:2000pz}). It is gratifying
to note that no truncation to the Dyson--Schwinger equations has to be applied
to arrive at this conclusion  if one assumes that the involved Green's
functions can be represented in the infrared by asymptotic expansions
\cite{Watson:2001yv,Lerche:2001}.  
In Landau gauge, a sufficient condition for the
Kugo--Ojima confinement criterion is that the nonperturbative ghost propagator
is more singular than a massless pole in the infrared \cite{Kugo:1995km}:
\begin{equation} 
D^{ab}_{G}(p) = - \delta^{ab} \frac {G(p^2)}{p^2} \, , \;\;
\mbox{with} \quad G(p^2) \stackrel {p^2 \to 0 } {\longrightarrow}\infty \, .
\label{GhProp} 
\end{equation} 
This behavior is also correlated to other aspects of Yang--Mills theories.
First, the Oehme--Zimmermann superconvergence relations~\cite{Oehme:1980ai} 
can be derived from Ward--Takahashi identities assuming the Kugo--Ojima
confinement criterion~\cite{Nishijima:1996ji}. These superconvergence relations
formalize a long known contradiction between asymptotic freedom and the
positivity of the spectral density for transverse gluons in the covariant
gauge.
Second, eq.\ (\ref{GhProp}) agrees with Zwanziger's horizon condition 
\cite{Zwanziger:1991gz,Zwanziger:2001kw}. This amounts to Gribov's prescription
to cut off the functional integral at the first Gribov horizon 
\cite{Gribov:1978wm}. Noting that this horizon is a convex hypersurface in
$A$-space that surrounds the origin \cite{Zwanziger:1982na} allows one to
conclude that the Dyson--Schwinger equations are not changed
\cite{Zwanziger:1991gz}. However, one has to note that this treatment of the
functional integral is related to the resolution of an ambiguity in the
solution of these equations \cite{vonSmekal:1997is,Zwanziger:2001kw}.
Nevertheless, supplementing the Faddeev--Popov quantization with this
additional constraint might be not sufficient to provide an exact solution of
the problem because there exist Gribov copies within the first Gribov horizon
\cite{vanBaal:1992zw}. 

As already stated, the obtained values for the infrared exponents of the gluon
and ghost propagators depend on the employed approximation for the
Dyson--Schwinger equations. Beyond the necessary truncation of this set of
integral equations in numerical calculations also some approximations for the
angular integrals have been used so far
\cite{vonSmekal:1997is,Atkinson:1998tu}. On the other hand, employing infrared
expansions (without using any angular approximation) which are strictly valid
only in the limit of vanishing momentum, $p^2\to 0$, 
\cite{Atkinson:1998tu,Watson:2001yv,Zwanziger:2001kw,Lerche:2001} yield also
some quite different values for the infrared critical exponent depending on the
truncation scheme. With respect to these analytical calculations the question
arises whether for every extracted value of the infrared exponent the
corresponding numerical solution exists also for finite values of momenta. As
will be detailed in this paper a tool to overcome  angular approximations is
the treatment of the Dyson--Schwinger equations on a compact Euclidean
four-manifold, in our case a four-torus. This allows us to answer the above
question: We will see that not every analytically extracted infrared exponent
can be matched to a numerical solution for non-vanishing momenta.

%This paper is organized as follows:  

To make this paper self-contained we will shortly summarize truncation schemes
for the gluon and ghost Dyson--Schwinger equations in flat (infinite) Euclidean
space-time, which have been solved recently, in the first two sections of Sect.\
II. In the following section we introduce  a novel truncation scheme. In
Sect.\ III we present the Dyson--Schwinger equations  formulated on the
momentum grid which is the dual space to the compact four-torus. In Sect.\ IV
we will present solutions done in otherwise exactly the same  approximation
scheme as previous solutions for flat space-time. The comparison to these
previous solutions allows to chose a suitable regularization and
renormalization procedure. (As in previous work we adopt  a modified MOM
scheme.)  In Sect.\ IV also the solutions without any angular approximations
will be discussed. The central result of this paper is: only one of two  
solutions allowed by the infrared analysis can be matched to a numerical
solution for non-vanishing momenta. {\it E.g.\/} for bare vertices this 
implies that the gluon propagator is only weakly infrared vanishing,
$D_{gl}(k^2) \propto (k^2)^{2\kappa -1}$, $\kappa =0.595\ldots$, and the ghost
propagator is infrared singular, $D_{gh}(k^2) \propto (k^2)^{-\kappa -1}$. In
Sect.\ V we present our conclusions. Furthermore,  in Appendix A we present 
our approximation to impose  
one-loop scaling of the gluon and ghost propagators in the ultraviolet.
In Appendix B we discuss the influence of gluon and ghost zero modes on the
solutions. The numerical methods to solve the gluon and ghost Dyson--Schwinger
equations in flat Euclidean space-time will be described in Appendix C.

\section{II. Gluon and ghost 
Dyson--Schwinger equations in flat Euclidean space-time}

\begin{figure}
  \centerline{ \epsfig{file=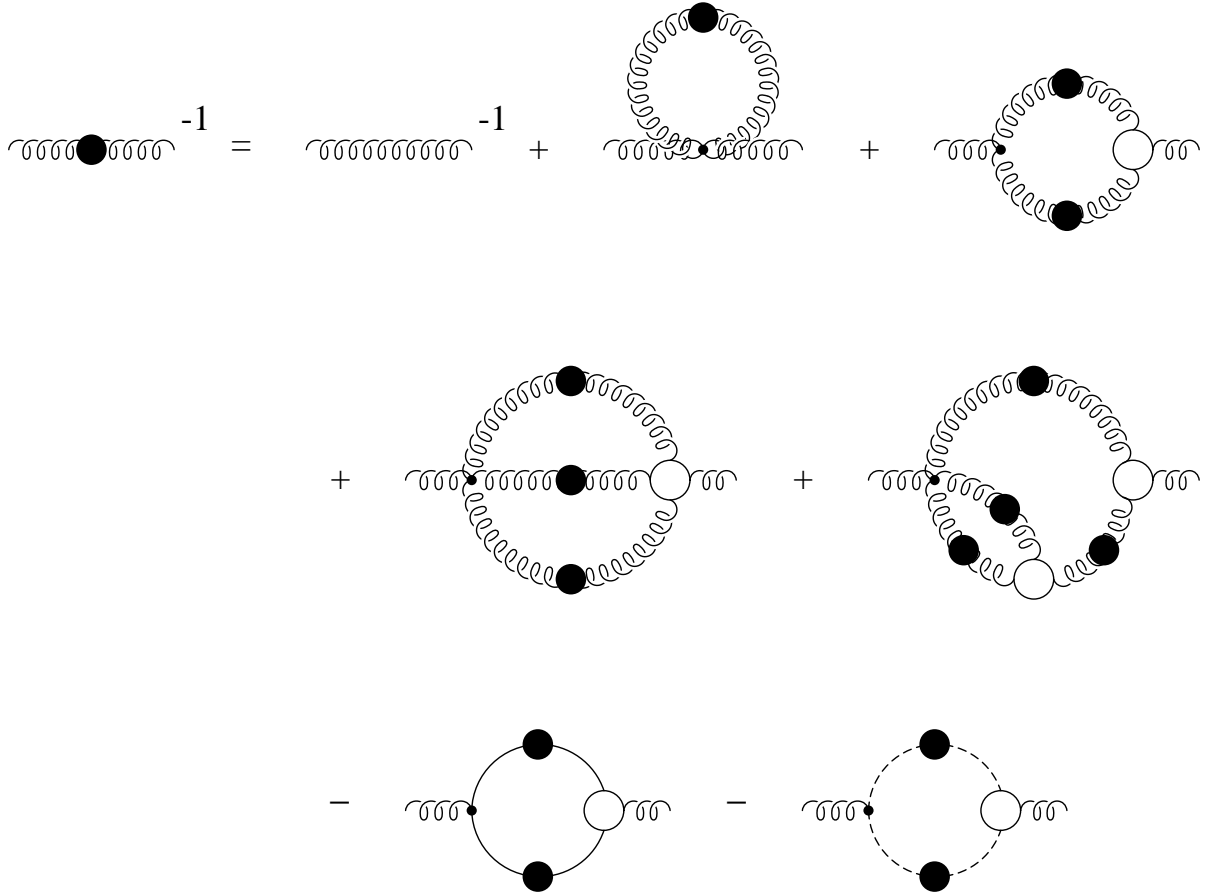,width=0.98\linewidth} }
  \vskip 5mm
  \caption{Diagrammatic representation of the gluon
           Dyson--Schwinger equation. The wiggly, dashed and solid
           lines represent the propagation of gluons, ghosts and quarks,
           respectively. A filled blob represents a full propagator and
           a circle indicates a one-particle irreducible vertex. }
  \label{Gluon}
\end{figure}

\begin{figure}
  \centerline{ \epsfig{file=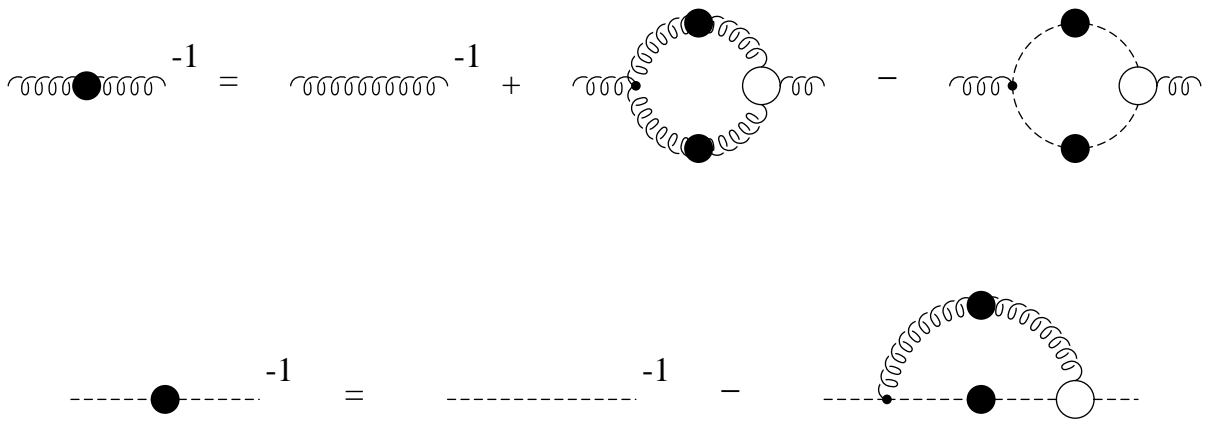,width=0.98\linewidth} }
  \vskip 3mm
  \caption{Diagrammatic representation of the truncated gluon and ghost 
  Dyson--Schwinger equations studied in this article. In the gluon 
  Dyson--Schwinger equation terms with four--gluon vertices and quarks
  have been dismissed.}
  \label{GluonGhost}
\end{figure}

In this section a short summary of previously employed truncation and 
approximation schemes for the coupled gluon and ghost Dyson--Schwinger
equations \cite{vonSmekal:1997is,Atkinson:1998tu} will be given first. We will
also provide the underlying formula for a new truncation scheme. All these
schemes include all diagrams in the ghost equation and neglect contributions
from the two-loop diagrams in the gluon equation, see Fig.\ \ref{Gluon}, where
the full gluon Dyson--Schwinger equation of QCD is represented
diagrammatically. In addition, as we will be only concerned with pure
Yang--Mills theory in this paper the quark loop will be neglected. The tadpole
term provides in Landau gauge only an (ultraviolet divergent) constant and will
drop out during renormalization anyhow. Thus, we will effectively study the
coupled system of equations as depicted in Fig.\ \ref{GluonGhost}.

The necessary trunction of the gluon Dyson--Schwinger equation immediately 
leads to a problem: The gluon polarization, which due to gauge symmetry would
be transverse to the gluon momentum in an exact calculation, acquires spurious
longitudinal terms. These terms are in general quadratically ultraviolet
divergent and thus highly ambiguous because they depend on the momentum routing
in the loop integral. In addition, a gauge invariant regularization scheme is
required to avoid these unphysical longitudinal terms.  Such schemes are,
however, hard to implement in Dyson--Schwinger studies, for the corresponding
use of dimensional regularization see {\it e.g.\/} refs.\
\cite{Gusynin:1999se,vonSmekal:1991fp}. 
An alternative unambiguous procedure is to isolate the part free of quadratic
ultraviolet divergences by contracting with the projector 
\begin{equation}
{\mathcal R}_{\mu\nu} (k) = \delta_{\mu\nu} - d \frac{k_\mu k_\nu}{k^2} =
\delta_{\mu\nu} - 4 \frac{k_\mu k_\nu}{k^2} \, , 
\label{Rproj}
\end{equation} 
which is constructed such that ${\mathcal R}_{\mu\nu} (k) \delta^{\mu\nu} =0$,
and therefore the ambiguous term proportional to $\delta_{\mu\nu}$ is projected
out \cite{Brown:1988bm}. Note that the use of this projector also removes 
the tadpole term.
As has become obvious recently \cite{Lerche:2001}
(see also ref.\ \cite{Hauck:1998sm} for a corresponding discussion in a much
simpler truncation scheme) the use of the projector (\ref{Rproj}) interferes 
with the infrared analysis of the coupled gluon-ghost system. For technical 
reasons we will employ a one-parameter family of projectors
\begin{equation}
{\mathcal P}^{(\zeta )}_{\mu\nu} (k) = \delta_{\mu\nu} - 
\zeta \frac{k_\mu k_\nu}{k^2} 
\, , 
\label{Paproj}
\end{equation} 
which allows us to interpolate continuously from the projector (\ref{Rproj}) to
the transversal one (with $\zeta =1$). Furthermore, we will also use the
general form of the quadratically ultraviolet divergent tadpole to remove
these unwanted ultraviolet divergencies.

Here we employ the conventions and notations of ref.\
\cite{Alkofer:2001wg}. As usual for Dyson--Schwinger studies all integrals are
formulated in Euclidean space-time. $Z(k^2)$ is the gluon renormalization
function defined via the gluon propagator in Landau gauge
\begin{equation}
 D^{ab}_{\mu\nu}(k) \, =\,  \delta^{ab} 
 \left(\delta_{\mu\nu} - \frac{k_\mu k_\nu}{k^2}
  \right) \,   \frac{Z(k^2)}{k^2} \; .
 \label{eq:Gluon-Prop}
\end{equation}
The deviation of $Z(k^2)$ from its tree-level value $Z\equiv 1$ provides a 
measure for renormalization of the gluon field due to the considered
interactions. The ghost renormalization function $G(k^2)$ is defined 
analogously via the ghost propagator, see eq.\ (\ref{GhProp}).

In Landau gauge the ghost-gluon vertex does not attribute an independent
ultraviolet divergence, {\it i.e.\/} one has $\widetilde Z_1 = 1$, 
\cite{Taylor:1971ff}. Therefore a truncation based on the tree-level form
for the ghost-gluon vertex function, $G_\mu (q,p) = iq_\mu $ is compatible
with the desired short distance behaviour of the solutions. This will be 
exploited in the following.

To proceed we will first consider the truncation schemes of refs.\
\cite{vonSmekal:1997is,Atkinson:1998tu} in the next two subsections.  Both
these truncation schemes employ only the projector $ {\mathcal R}_{\mu\nu}$
(\ref{Rproj}) in the gluon equation. The main difference between these
truncation schemes consists in the treatment of the three-point functions.
Whereas in ref.\ \cite{vonSmekal:1997is} the form of  ghost-gluon and
three-gluon vertex function has been related to the gluon and ghost
renormalization functions using Slavnov--Taylor identities (and then the
resulting system has been solved self-consistently), in ref.\
\cite{Atkinson:1998tu} bare three-point functions have been used. Amazingly,
though, both schemes provide results with identical qualitative infrared
behavior: the gluon propagator vanishes in the infrared, the ghost propagator
is highly singluar there, and the strong running coupling (which can be related
to the gluon and ghost renormalization functions using the specific form of the
ghost-gluon vertex in Landau gauge \cite{vonSmekal:1997is}) has an infrared fix
point. Because this infrared behavior is determined by the interplay between
the ghost loop in the gluon equation (the gluon loop being subleading in the
infrared) and the ghost equation such a scenario is also found in the
ghost-loop only approximation \cite{Atkinson:1998tu}. We will exploit these two
recent truncation schemes in the course of this article for various tests of
our method.

\subsection{A. The dressed vertex truncation including the gluon loop}

In  ref.\  \cite{vonSmekal:1997is} an approximation scheme for the
longitudinal parts of ghost-gluon  and 3-gluon vertex functions has been
employed which ensures consistency at the level of 
one-particle Greens functions, {\it i.e.\/} propagators. The detailed form 
of the vertex functions can be found in ref.~\cite{vonSmekal:1997is}. 
Here we provide directly the ghost,
\begin{equation}
  \frac{1}{G(k^2)} \, =\,  \widetilde{Z}_3 -  g^2 N_c \int {d^4q\over
  (2\pi)^4}  \, \biggl( k {\cal P}^{(1)}(p) q \biggr) \,
  \frac{Z(p^2) G(q^2)}{k^2\, p^2\, q^2} \,
  \label{ghDSE1}
\end{equation}
\[ \hskip 50mm
  \times \left( \frac{G(p^2)}{G(q^2)} + \frac{G(p^2)}{G(k^2)} - 1 \right) \; ,
  \quad p = k - q\; ,
\]
and the gluon
\begin{eqnarray}
  \frac{1}{Z(k^2)} &=&
     Z_3 \,-\, Z_1\frac{ g^2 N_c}{6} \int \frac{d^4q}{(2\pi)^4} \, \left\{
       N_1(p^2,q^2;k^2)  \, \frac{Z(p^2)G(p^2) Z(q^2)G(q^2)}{Z(k^2)G^2(k^2)}
     \right. \nonumber \\
     &&\hskip -1.5cm \left. + \,
        N_2(p^2,q^2;k^2) \, \frac{Z(p^2)G(p^2)}{G(q^2)} \, +
        N_2(q^2,p^2;k^2) \, \frac{Z(q^2)G(q^2)}{G(p^2)}\,
\right\} \,    \frac{G(k^2)}{k^2\, p^2\, q^2}  
%\nonumber \\
\label{ZDSE} \\
    && \hskip -2cm +  \frac{ g^2N_c}{3} \int \frac{d^4q}{(2\pi)^4}
       \Bigg\{ \Bigl(q {\cal R}(k) q\Bigr)
\Bigl( G(k^2) G(p^2) - G(q^2) G(p^2) \Bigr)
% \nonumber \\
%    &&\hskip 2cm
                 - \Bigl( q {\cal R}(k) p \Bigr)  G(k^2) G(q^2) \Bigg\}
                \frac{1}{k^2\, p^2\, q^2}  \; ,   \nonumber %\label{ZDSE}
\end{eqnarray}
equations, respectively. The functions $N_1(x,y;z) = N_1(y,x;z)$ and 
$N_2(x,y;z)$ are given in appendix C of ref.\ \cite{vonSmekal:1997is}.

In eq.\ (\ref{ghDSE1}) we have already exploited the identity  $\widetilde Z_1
= 1$. This leaves the gluon and ghost field renormalization constants $Z_3$ and
$\widetilde Z_3$ as well as the gluon vertex renormalization constant $Z_1$ to
be determined  correspondingly to the employed truncation. Note that these
constants depend on the ultraviolet cutoff $\Lambda$ and the renormalization
scale $\mu$.

In ref.\ \cite{vonSmekal:1997is} different angular approximations  for $q^2 >
k^2 $  and  for $q^2 < k^2$ have been employed. In the latter case $G(p^2) =
G((k-q)^2) \to G(k^2)$ and $Z(p^2) \to Z(k^2) $ have been set which obviously
preserves the limit $q^2 \to 0$ of the integrand.
With this approximation one obtains from eq.\ (\ref{ghDSE1}) 
upon angular integration
\begin{eqnarray}
 \frac{1}{G(k^2)} &=&  \widetilde{Z}_3 - \frac{g^2}{16\pi^2} \frac{3 N_c}{4} \left\{
\int_0^{k^2} \, \frac{dq^2}{k^2} \frac{q^2}{k^2} \, Z(k^2) G(k^2)
\, + \,\int_{k^2}^{\Lambda^2} \frac{dq^2}{q^2} \,  Z(q^2) G(q^2) \right\} \nonumber \\
&=&  \widetilde Z_3 -  \frac{g^2}{16\pi^2} \frac{3 N_c}{4}  \left( \frac{1}{2} \,
Z(k^2) G(k^2) \, + \,\int_{k^2}^{\Lambda^2} \frac{dq^2}{q^2} \,  Z(q^2) G(q^2)
\right) \; ,
  \label{odGDSE}
\end{eqnarray}
where we introduced an $O(4)$--invariant momentum cutoff $\Lambda$ to account
for the logarithmic ultraviolet divergence, which will have to be absorbed by
the renormalization constants.
With some further assumptions the angular approximation for the gluon equation 
(\ref{ZDSE}) yields
\begin{eqnarray}
  \frac{1}{Z(k^2)}
     &=& Z_3 + Z_1 \frac{g^2}{16\pi^2} \frac{N_c}{3}
         \left\{ \int_{0}^{k^2} \frac{dq^2}{k^2}
         \left( \frac{7}{2}\frac{q^4}{k^4}
         - \frac{17}{2}\frac{q^2}{k^2}
         - \frac{9}{8} \right) Z(q^2) G(q^2) \right. \nonumber \\
     && + \, \left. \int_{k^2}^{\Lambda^2} \frac{dq^2}{q^2} \left(
            \frac{7}{8} \frac{k^2}{q^2} - 7 \right) Z(q^2) G(q^2)
          \right\} \label{odZDSE} \\
     && \hskip -2cm
+ \, \frac{g^2}{16\pi^2} \frac{N_c}{3} \left\{ \int_{0}^{k^2}
         \frac{dq^2}{k^2}
         \frac{3}{2} \frac{q^2}{k^2} G(k^2) G(q^2) - \frac{1}{3} G^2(k^2)
         + \frac{1}{2} \int_{k^2}^{\Lambda^2} \frac{dq^2}{q^2}
         G^2(q^2) \right\} \; . \nonumber
\end{eqnarray}

In the infrared the solutions $Z(x)$ and $G(x)$ behave power-like,
\begin{equation}
Z(x) \propto x^{2\kappa} , \quad G(x) \propto x^{-\kappa} .
\label{power}
\end{equation}
In this truncation scheme one obtains  $\kappa \approx 0.92$
\cite{vonSmekal:1997is}. To solve the coupled system for all momenta the power
laws, eq. (\ref{power}), are used to perform the integrals from $y=0$ to an
infrared point $y=\epsilon$ analytically, while the remaining part of the
integrals is done with the help of numerical routines, see {\it e.g.\/}
ref.\ \cite{Hauck:1998sm}.  

The non-perturbative subtraction scheme of ref.\ \cite{vonSmekal:1997is}
implies a strong running coupling with infrared fixed point.
Starting again from the non-renormalization of the ghost-gluon vertex
\begin{equation}
  \widetilde{Z}_1 \, = \, Z_g Z_3^{1/2} \widetilde{Z}_3 \, = \, 1 \; ,
  \label{wtZ1}
\end{equation}
one can readily show that the product $g^2 Z(\mu^2) G^2(\mu^2)$ is
renormalization group invariant. Therefore, in absence of any dimensionful
parameter, this (dimensionless) product is a function of the running coupling
$\bar g$ only. Analysing this renormalization group invariant product more
closely one concludes that it is identical to the running coupling $\bar{g}^2
(\mu^2)$. As the infrared powers in the product $ZG^2$ cancel exactly 
the running coupling is a finite constant for $\mu^2=0$.

\subsection{B. The bare vertex ghost-loop only truncation}

Substituting the tree-level ghost-gluon vertex for the dressed one and
neglecting the gluon loop the coupled system of equations 
(\ref{ghDSE1},\ref{ZDSE})  reads 
\begin{eqnarray} 
\frac 1 {G(k^2)} &=& \widetilde Z_3 - g^2 N_c \widetilde Z_1
\int \frac{d^4q}{(2\pi)^4} \frac{k^2q^2-(k\cdot q)^2}{k^2q^2(k-q)^4}
G(q^2)Z((k-q)^2) ,
\label{ghost} \\
\frac 1 {Z(k^2)} &=& Z_3 + g^2 \frac {N_c}{3} 
\widetilde Z_1 \int \frac{d^4q}{(2\pi)^4}
\frac {k^2(k^2+q^2+(k-q)^2)+4q^2(k-q)^2-2(k-q)^4-2q^4} {2k^4q^2(k-q)^2} 
\nonumber \\
&& \qquad \qquad \qquad \qquad \qquad \qquad \qquad \qquad 
\qquad \qquad \qquad \times G(q^2)G((k-q)^2)  .
\label{gluon1} 
\end{eqnarray} 

To simplify notation we introduce the abbreviations $x:=k^2$, $y:=q^2$, $s:=\mu^2$
and $L:=\Lambda^2$. In the angular approximation of ref.\
\cite{Atkinson:1998tu} (`ymax approximation`) where $Z({\rm min}(p^2,q^2))$ and
$G({\rm min}(p^2,q^2))$ are substituted for $Z((k-q)^2)$ and $G((k-q)^2)$ eqs.\
(\ref{ghost},\ref{gluon1}) are simplified to
\begin{eqnarray}
\frac 1 {G(x)} &=& \widetilde Z_3(s,L) - \frac 9 4 \frac {g^2 N_c}{48\pi^2}
\left( Z(x) \int_0^x \frac{dy}{x} \frac{y}{x} G(y)
+ \int_x^L \frac{dy}{y} Z(y) G(y) \right) ,
\label{ghAB}\\
\frac 1 {Z(x)} &=& Z_3(s,L) + \frac {g^2 N_c}{48\pi^2}
\left( G(x) \int_0^x \frac{dy}{x} \left( -\frac{y^2}{x^2} + \frac{3y}{2x} 
\right) G(y) + \int_x^L \frac{dy}{2y} G^2(y)\right) .
\label{glAB}
\end{eqnarray} 
Imposing as renormalization conditions $Z(s)=G(s)=1$ to determine $Z_3$ and
$\widetilde Z_3$ these equations may be solved numerically for a given 
(O(4) invariant) cutoff.
As a matter of fact, we use subtracted finite equations in our numerical
procedure, see Appendix C for details. 

In the infrared, also the solutions of eqs.\ (\ref{ghAB},\ref{glAB})  behave
power-like, {\it c.f.\/} eq.\ (\ref{power}), with  $\kappa \approx 0.77$
\cite{Atkinson:1998tu}. In the same truncation but with no angle approximation
employed a solution $\kappa = 1$ has been extracted for the infrared
behaviour. However, it will be
explained below that we could not find a numerical  solution for non-vanishing
momenta connected to the $\kappa = 1$ infrared behaviour.

\subsection{C. The bare vertex truncation including the gluon loop}

In this subsection we will detail a novel truncation scheme which employs bare
vertices. Nevertheless it will be constructed such that it reproduces  the
correct perturbative limit for large momenta. To analyse the gluon loop we will
use the class of projectors 
${\mathcal P}^{(\zeta)}_{\mu\nu} $ (\ref{Paproj}). 
A smooth interpolation between the Brown--Pennington projector ($\zeta=4$) and
the transverse one ($\zeta=1$) will be helpful in the analysis of unphysical 
quadratic ultraviolet divergencies for $\zeta \ne 4$ in the gluon equation.
Their careful removal is essential for a numerical solution retaining the 
infrared behaviour of the solutions. 

The coupled equations for the ghost and gluon dressing functions using bare
vertices read as follows (The ghost equation is, of course, identical
to eq.\ (\ref{ghost}), it is only repeated for a coherent representation.):
\begin{eqnarray} 
\frac{1}{G(x)} &=& Z_3 - g^2N_c \int \frac{d^4q}{(2 \pi)^4}
\frac{K(x,y,z)}{xy}
G(y) Z(z) \; , \label{ghostbare} \\ 
\frac{1}{Z(x)} &=& \tilde{Z}_3 + g^2\frac{N_c}{3} 
\int \frac{d^4q}{(2 \pi)^4} \frac{M(x,y,z)}{xy} G(y) G(z) + Z_1
 g^2 \frac{N_c}{3} \int \frac{d^4q}{(2 \pi)^4} 
\frac{Q(x,y,z)}{xy} Z(y) Z(z) \; .\nonumber\\
\label{gluonbare} 
\end{eqnarray} 
The kernels ordered with respect to powers of $z:=p^2=(k-q)^2$ have the form:
\begin{eqnarray}
K(x,y,z) &=& \frac{1}{z^2}\left(-\frac{(x-y)^2}{4}\right) + 
\frac{1}{z}\left(\frac{x+y}{2}\right)-\frac{1}{4}\\
M(x,y,z) &=& \frac{1}{z} \left( \frac{\zeta-2}{4}x + 
\frac{y}{2} - \frac{\zeta}{4}\frac{y^2}{x}\right)
+\frac{1}{2} + \frac{\zeta}{2}\frac{y}{x} - \frac{\zeta}{4}\frac{z}{x} \\
Q(x,y,z) &=& \frac{1}{z^2} 
\left( \frac{1}{8}\frac{x^3}{y} + x^2 -\frac{19-\zeta}{8}xy + 
\frac{5-\zeta}{4}y^2
+\frac{\zeta}{8}\frac{y^3}{x} \right)\nonumber\\
&& +\frac{1}{z} \left( \frac{x^2}{y} - \frac{15+\zeta}{4}x-
\frac{17-\zeta}{4}y+\zeta\frac{y^2}{x}\right)\nonumber\\
&& - \left( \frac{19-\zeta}{8}\frac{x}{y}+\frac{17-\zeta}{4}+
\frac{9\zeta}{4}\frac{y}{x} \right) \nonumber\\
&& + z\left(\frac{\zeta}{x}+\frac{5-\zeta}{4y}\right) + z^2\frac{\zeta}{8xy}
\label{new_kernels}
\end{eqnarray}
It is straigtforward to verify that for $\zeta=4$ the kernel $M(x,y,z)$
is identical to the kernel in eq.\ (\ref{gluon1}). 

The quadratic ultraviolet divergencies of the integrals are most easily
discovered approximating the angular integrals as done in the previous
subsections (note that we will use these approximations only for an analysis
of the ultraviolet behaviour) and introducing an Euclidean sharp cutoff. 
Displaying only the ultraviolet divergent integrals
\begin{equation}
\frac{g^2N_c}{48 \pi^2} \int_x^{L} \frac{dy}{x} \left( 
\left(\frac{4-\zeta}{4} + \frac{\zeta-2}{4}\frac{x}{y}
\right) G^2(y)
+ Z_1  \left(\frac{3\zeta-12}{2} - 
\frac{\zeta+24}{4}\frac{x}{y}
+\frac{7}{8}\frac{x^2}{y^2}\right) Z^2(y) \right)
\end{equation}
one sees that for $\zeta=4$ the terms independent of the integration momentum
$y$ vanish. Thus these integrals are then only logarithmically ultraviolet
divergent as could be expected on the basis of the results summarized in the
previous subsections. Of course, the quadratic ultraviolet divergencies are
artefacts of the employed truncation. Due to gauge invariance they would cancel
against the tadpole and similar divergencies in the two--loop terms. The
calculation of the latter being beyond the scope of this paper we will simply
subtract these divergent terms. This cannot be done straightforwardly at the
level of integrands: Such a procedure would disturb the infrared properties of
the Dyson--Schwinger equations. As we anticipate from previous studies and
analytic work \cite{Zwanziger:2001kw,Lerche:2001} that the ghost loop is the
leading contribution in the infrared the natural place to subtract the
quadratically ultraviolet divergent constant is the gluon loop. We do this 
by employing the substitution 
\begin{equation}
Q(x,y,z) \to Q^\prime (x,y,z) = Q(x,y,z) + \frac 5 4 (4-\zeta )
\end{equation}
in eq.\ (\ref{gluonbare}). At the first sight, due to the presence of the
prefactor $Z_1$ in the gluon loop, this seems not to be sufficient to remove 
also the quadratic ultraviolet divergence of this ghost loop. However, note that
in the next step we will enforce consistency of the logarithmic divergencies 
which entails then cancelation of the quadratic divergencies if $Q^\prime $
is employed in the gluon loop, see below.

To achieve the correct one-loop scaling in the ultraviolet we will adopt a
similar treatment to the one of ref.\ \cite{vonSmekal:1997is}. Please note that 
within the presented class of truncation schemes it is impossible to fulfill
both, correct one-loop scaling and the Slavnov--Taylor identity
$Z_1=Z_3/\tilde{Z}_3$.  For large Euclidean momenta and to one loop the 
behavior of the propagator functions can be described as 
\begin{eqnarray}
Z(x) &=& Z(s) \left[ \omega \log\left(\frac{x}{s}\right)+1 \right]^\gamma  \; ,
\label{gluon_uv}\\
G(x) &=& G(s) \left[ \omega \log\left(\frac{x}{s}\right)+1 \right]^\delta  \; .
\label{ghost_uv}
\end{eqnarray}
$Z(s)$ and $G(s)$ denote the value of the dressing functions at some
renormalization point $s:=\mu^2$, $\gamma$ and $\delta$ are the respective 
anomalous dimensions. To one loop one has $\delta = - 9/44$ and $\gamma = - 1
-2\delta$ for arbitrary number of colors $N_c$ and no quarks, $N_f=0$
\cite{Muta:1998vi}. Furthermore, $\omega = {11N_c}\alpha(s)/{12\pi}$.

Employing these expressions in the ghost equation  (\ref{ghostbare})
and approximating the angular integrals as done previously one obtains to the
order of the leading logarithms
\begin{eqnarray}
G^{-1}(s)  \left[ \omega \log\left(\frac{x}{s}\right)+1 \right]^{-\delta} &=&
\tilde{Z}_3 - \frac{9 g^2 Z(s) G(s)}{64 \pi^2 \omega(\gamma+\delta+1)} 
\times \nonumber\\
&&\left\{
\left[ \omega \log\left(\frac{L}{s}\right)+1 \right]^{\gamma+\delta+1} - 
\left[ \omega \log\left(\frac{x}{s}\right)+1 \right]^{\gamma+\delta+1} \right\}
\end{eqnarray}
where the abbreviation $L:=\Lambda^2$ has been used again. The renormalization
constant $\tilde{Z}_3(L,s)$ cancels the cutoff dependence of the first term in the
bracket. Thus, the power and the prefactor of the second term have to match with
the left hand side of this equation. This leads to two conditions:
\begin{eqnarray}
\gamma+2\delta+1 &=& 0 \; , \label{gd1}\\
\frac{9}{2 \omega(\gamma+1)} \frac{g^2}{16 \pi^2}Z(s) G^2(s) &=& 1 \; .
\label{go1}
\end{eqnarray}
Eq.\ (\ref{gd1}) is of course nothing else but consistency of the ghost
equation with one-loop scaling.

Plugging the ultraviolet behavior (\ref{gluon_uv}) and (\ref{ghost_uv})
into the gluon equation (\ref{gluonbare}) and keeping the leading order
logarithms leads to
\goodbreak
\begin{eqnarray}
Z^{-1}(s) \left[ \omega \log\left(\frac{x}{\mu}\right)+1 \right]^{-\gamma} &=&
Z_3 - \frac{g^2 G^2(s)}{32 \pi^2 \omega \gamma} \times \nonumber \\
&&\left\{
\left[ \omega \log\left(\frac{L}{s}\right)+1 \right]^{-\gamma} -
\left[ \omega \log\left(\frac{x}{s}\right)+1 \right]^{-\gamma} \right\} 
\nonumber \\
 &&- \frac{7g^2Z(s)G(s)}{16 \pi^2} \int_x^L dy Z_1
\left[\omega \log\left(\frac{y}{s}\right)+1 \right]^{\gamma+\delta} \; .
\label{gluoneq_uv}
\end{eqnarray}
Here we see that due to the employed truncation we have to give up either one-loop
scaling or the Slavnov--Taylor identity $Z_1=Z_3/\tilde{Z}_3$. Note, however,
that $Z_3(L,s)$ has the correct cutoff dependence to make this equation finite
with finite $Z_1(L,s)$. Nevertheless, in order to get the correct leading log 
contribution from the gluon loop the integrand in the last line
should be proportional to 
$\left[\omega \log\left(\frac{y}{\mu}\right)+1 \right]^{-\gamma-1}$.
This not only contradicts the identity $Z_1=Z_3/\tilde{Z}_3$ but also requires
the renormalization {\em constant} $Z_1(L,s)$ to acquire a momentum dependence
if correct one-loop scaling is to be enforced. Following case c) described
in chapter 6 of ref.\ \cite{vonSmekal:1997is} we demand therefore
\begin{equation} 
Z_1(L,s) \to {\cal Z}_1(x,y,z;s,L)  \sim 
\left[\omega \log\left(\frac{y}{\mu}\right)+1 \right]^{-\delta-2\gamma-1} \; .
\end{equation} 
Generalizing the approach of  ref.\ \cite{vonSmekal:1997is} we note that the 
following two-parameter ansatz
\begin{equation} 
{\cal Z}_1(x,y,z;s,L) = 
\frac{G(y)^{(1-a/\delta-2a)}}{Z(y)^{(1+a)}}
\frac{G(z)^{(1-b/\delta-2b)}}{Z(z)^{(1+b)}}
\label{Z1_ansatz}
\end{equation}
with arbitrary $a$ and $b$ satisfies the required proportionality.
This can be straightforwardly verified with the help of expression (\ref{gd1})
obtained from the ghost equation. In addition ansatz (\ref{Z1_ansatz}) ensures
the cancelation of quadratic ultraviolet divergencies as discussed above also
if $\zeta\not = 4$. 

Carrying out the remaining integral in (\ref{gluoneq_uv}) and matching powers 
one obtains
\begin{eqnarray}
\gamma+2\delta+1 &=& 0 \; , \label{gd2}\\
\frac{1}{2\gamma \omega}\frac{g^2}{16 \pi^2}Z(s) G^2(s) -
\frac{7}{\gamma \omega} \frac{g^2}{16 \pi^2}
{Z(s)^{1-a-b}}{G(s)^{2-a/\delta-b/\delta-2a-2b}} &=& 1 \; .
\end{eqnarray}
First of all, we reproduce correct one-loop scaling also from the gluon equation
as can be seen from the equivalence of eqs.\ (\ref{gd2}) and (\ref{gd1}).
Second, with the perturbative renormalization conditions $Z(s)=G(s)=1$ 
which are possible and appropriate at a large Euclidean renormalization point
$\mu$ one obtains together with eq.\ (\ref{go1}) the correct anomalous
dimensions $\gamma=-13/22$ and $\delta=-9/44$. 

A reasonable choice of parameters is of course one which keeps ${\cal Z}_1$ as
weakly dependent as possible on the momenta $y$ and $z$,
{\it c.f.}~Fig.~\ref{z1.dat} in Appendix A. Note that the choice
$a=b=0$ corresponds to the truncation scheme of \cite{Bloch:2001wz} 
whereas $a=3\delta, b=0$ together with the appropriate vertex dressings 
reproduces case c) of ref.\ \cite{vonSmekal:1997is}.
The infrared behavior of the gluon loop in the gluon equation depends
strongly on $a$ and $b$. {\it E.g.\/} for $b=0$ one can distinguish three
cases: For $a<0$ the gluon loop is subleading in the infrared, for
$a=0$ as in ref.\ \cite{Bloch:2001wz} the gluon loop produces the same power as
the ghost loop, for $a>0$ the gluon loop becomes the leading term in the
infrared. In the latter case we did not find a  solution to the coupled
gluon-ghost system.
In Appendix A we will demonstrate that $a=b=3\delta$ minimizes the momentum 
dependence of ${\cal Z}_1$. Thus we will use these values except stated
otherwise explicitely.

The infrared behavior of the propagator functions in this truncation scheme 
and for the transverse projector ($\zeta=1$) has been determined very recently 
\cite{Zwanziger:2001kw,Lerche:2001}. It can be shown straightforwardly that 
the relation between ghost and gluon infrared behavior is again as in eq.\
(\ref{power}). This is expected also on general grounds
\cite{Watson:2001yv}.   Due to this one can immediately use the ans\"atze
\begin{equation}  
Z(x) \to A x^{2\kappa } \; , \quad G(x) \to B x^{-\kappa }
\quad {\rm for} \quad x \to 0 \; . 
\label{ZABG}
\end{equation} 
%We will furthermore assume that the gluon loop is subleading, {\it i.e.\/}
%we will not consider the case $a=b=0$. 
As is explained in detail in ref.\
\cite{Lerche:2001} the renormalization constants $Z_3$ and $\tilde{Z}_3$ can be
dropped for very small momenta $x$: They are either subleading in the infrared
(gluon equation) or have to be zero when the renormalization takes place at
$\mu=0$ (ghost equation). The remaining infrared integrals can be evaluated
using the formula \cite{Lerche:2001}
\begin{equation}
\int d^4q \: y^a z^b = \pi^2 x^{2+a+b} \frac{\Gamma(2+a)\Gamma(2+b)\Gamma(-a-b-2)}
{\Gamma(-a)\Gamma(-b)\Gamma(4+a+b)} \; ,
\label{irintegral}
\end{equation}
where again $x=k^2$, $y=q^2$ and $z=(k-q)^2$.
From the resulting two conditions
\begin{eqnarray} 
\frac{1}{18}
\frac{(2+\kappa)(1+\kappa)}{(3-2\kappa)}
 &=& \frac{\Gamma^2(2-\kappa)\Gamma(2\kappa)}{\Gamma(4-2\kappa) \Gamma^2(1+\kappa)}
\: \frac{g^2N_c}{48 \pi^2}{AB^2} \; ,
\label{kappa1}\\
\frac{4\kappa-2}{4\zeta \kappa - 4\kappa + 6 - 3\zeta}
&=& \frac{\Gamma^2(2-\kappa)\Gamma(2\kappa)}{\Gamma(4-2\kappa) \Gamma^2(1+\kappa)}
\:\frac{g^2N_c}{48 \pi^2}{AB^2} \; ,
\label{kappa2}
\end{eqnarray} 
%%%%%%%%%%%%%%%%%%%%%%%%%%%%%%%%%%%%%%%%%%%%%%%%%%%%%%%%%%%%%%%%%%%%%%%%%%%%%%%%%%%%%
\begin{figure}
\centerline{
\epsfig{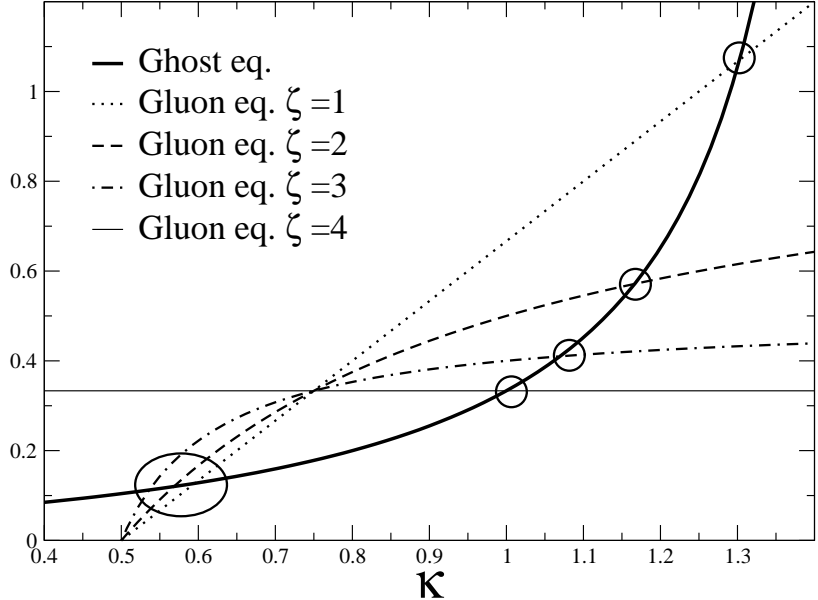}
}
\caption{\label{kappa.dat}
Here the graphical solution to equations (\ref{kappa1}) and (\ref{kappa2}) is shown. 
The thick line represents the
left hand side of equation (\ref{kappa1}), whereas the other four curves depict 
the left hand side of equation 
(\ref{kappa2}) for different values of the parameter $\zeta$. The ellipse marks the 
bulk of solutions between
$\kappa=0.5$ and $\kappa=0.6$ for different $\zeta$, whereas the circles show the 
movement of the solution $\kappa=1.3$ 
for a transverse projector to $\kappa=1$ for the Brown-Pennington case, $\zeta=4$.
}
\end{figure}
%%%%%%%%%%%%%%%%%%%%%%%%%%%%%%%%%%%%%%%%%%%%%%%%%%%%%%%%%%%%%%%%%%%%%%%%%%%%%%%%%%%%%
one obtains the determining relation for $\kappa$ by equating both left hand
sides, see Fig.~\ref{kappa.dat}. For the Brown--Pennington projector, 
{\it i.e.} $\zeta=4$, one then
finds the known solution $\kappa =1 $ \cite{Atkinson:1998tu}. However,
as can be seen immediately the left hand side of the second equation 
possesses a zero for $\kappa=1/2$ which is canceled by a pole only for $\zeta=4$. 
Lowering $\zeta$ only slightly a further solution with $\kappa$
slightly larger than 0.5 exists. For the transverse projector, {\it i.e.}
$\zeta=1$, this latter solution becomes $\kappa = 0.59...$ in accordance with
refs.\ \cite{Zwanziger:2001kw,Lerche:2001}. Also the solution $\kappa =1 $
changes continuously when lowering $\zeta$. The corresponding $\kappa$  are
then all larger than 1 and contradict the masslessness condition, see chapter 5
of ref.\  \cite{Alkofer:2001wg} for a discussion of this condition. The main
result of this paper is that the infrared behavior $\kappa \approx 0.5$ matches
to a numerical solution whereas no numerical solution could be found with the
infrared behavior $\kappa =1 $.

The renormalization group analysis for the running coupling given 
in ref.~\cite{vonSmekal:1997is} 
certainly applies here as well. The renormalization group
invariant expression for the running coupling is therefore given by
\begin{equation}
\alpha(x) =  \frac{g^2}{4 \pi} Z(x) G^2(x) \; .
\end{equation}
As can be seen directly from equations (\ref{kappa1},\ref{kappa2}),  in the
here presented truncations the product $N_c g^2 AB^2$ is constant for a given
$\kappa$. With $\alpha(0) = {g^2}AB^2/{4 \pi}$ one concludes immediately that
$\alpha(x)$ is proportional to $N_c^{-1}$.  Furthermore, the ghost and gluon
dressing functions $Z(x)$ and $G(x)$ are independent of the number of colors:
$N_c$ enters the Dyson-Schwinger equations only in the combination 
$g^2 N_c$ at our
level of truncation. From the solution $\kappa = 0.5953$ of the infrared
analysis with the transverse projector $\zeta=1$ one determines the infrared
fixed point of the running coupling to be $\alpha(0) = 2.972$ for $N_c=3$.

\goodbreak

\section{III. Gluon and ghost Dyson--Schwinger equations on a four-torus}

There are three central aims connected to the investigation of the
Dyson--Schwinger equations on a four-torus. The first one is purely technical:
This allows to study finite volume effects also in Dyson--Schwinger approach.
Monte-Carlo simulations on a lattice necessarily have to be done in a finite
volume. Therefore in the latter kind of approach infrared properties are only
accessible by extrapolations to an infinite volume where the avalaible data are
gained on several different volumes which, due to limitations in computer time,
do only differ at best by one order of magnitude. We will see in the present
Dyson--Schwinger approach that available volumes cover several orders of
magnitude. And more importantly, in several truncations and approximations one
can compare to the results obtained in an infinite volume.

The second issue is the solution of the (truncated) Dyson--Schwinger equations 
without any angular approximations. We will detail below why and how this can be
achieved on the momentum grid dual to the four-torus. This will also lead us
to the main result of this paper discussed in the following section.

The third aim relates to the possibility of topological obstructions on a
compact manifold. It is well known {\it e.g.\/} that a four-torus allows for a
non-vanishing Pontryagin index. We hope that we will be able to describe the
solutions of Dyson--Schwinger equations with twisted boundary conditions
\cite{'tHooft:1979uj} in a
subsequent publication. And choosing an asymmetric four-torus might allow the
introduction of a non-vanishing temperature in a relatively simple way.

\subsection{A. Finite volume effects}

From a technical point of view using as underlying manifold a four-torus or
choosing periodic boundary conditions on a hypercube is identical. Note that
the definition of the Faddeev--Popov operator necessitates periodic boundary
conditions for ghosts instead of their Grassmannian nature, see {\it e.g.\/}
ref.\ \cite{Reinhardt:1996fs}. With $L$ being the
length in every direction of the hypercube the four-dimensional momentum
integral has to be substituted by a sum over four indices,
\begin{equation}
\int \frac {d^4q}{(2\pi )^4} \to \frac {1}{L^4} \sum _{j_1,j_2,j_3,j_4} \; .
\label{DefineSum}
\end{equation}
On the other hand, the quantities of interest, the gluon and ghost
renormalization function, $Z(k^2)$ and $G(k^2)$, resp., do only depend on the
O(4) invariant squared momenta as long as all directions are treated on an
equal footing on the torus. This suggests to relabel the points on the
momentum grid not according to a Cartesian but a hypersherical coordinate 
system,
\begin{equation} 
\frac {1}{L^4} \sum _{j_1,j_2,j_3,j_4} = \frac {1}{L^4} \sum _{j,l} \; ,
\label{ShericalSum} 
\end{equation}
where the index $j$ numbers the hyperspheres $q^2={const}$. The index $l$, which numbers
the grid points on each hypersphere respectively, will be dropped in the following. 

In the integrals to be discretized there appear three momenta, the external
momentum, labeled $k$, the loop momentum $q$ and for the
second propagator in the loop $p=k-q$. We will use the following notation:
\begin{eqnarray}
x:= k^2 \qquad &{\rm with}& \qquad x_i \in {\rm hypersphere} \; i \; ,
\nonumber\\
y:= q^2 \qquad &{\rm with}& \qquad y_j \in {\rm hypersphere} \; j \; ,
\nonumber\\
z:= p^2=(k-q)^2 \qquad &{\rm with}& \qquad z_n \in {\rm hypersphere} \; n \; .
\label{xyzDef}
\end{eqnarray} 
On the hypercubic momentum grid dual to the four-torus the momentum $p=k-q$
is located on the grid for every pair of grid momenta $k$ and $q$ as can
be seen from elementary vector operations (or from the analogue of cubic
lattices in solid state physics).

\begin{figure}
\begin{center}
\epsfig{file=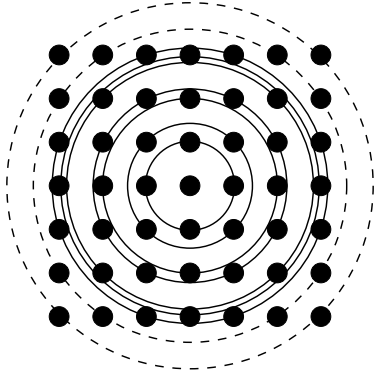,width=0.3\linewidth}
\end{center}
\caption{\label{Latt}
Scetch of the momentum grid dual to the four-torus and the 
summation over complete hyperspheres indicated by fully drawn
circles. The hyperspheres depicted by dashed lines are not complete
due to the numerical ultraviolet cutoff.}
\end{figure}

Note that we have introduced an O(4) invariant cutoff $\Lambda$ into eqs.\
(\ref{odGDSE},\ref{odZDSE}). As we will see below in more detail a
corresponding regularization of the sums over grid momenta is required.
In a first step we set up a momentum grid with $j_1,j_2,j_3,j_4=-N, \ldots ,
0, \ldots N$.
The hypercube property of this lattice seem to suggest in the first
place to cut off the sums such that over complete hypercubes is summed. Such a
method, however, breaks O(4) invariance and introduces sizeable numerical 
errors. (Note that in lattice Monte-Carlo simulations the analysis of the
resulting data for O(4) invariance necessitates special kinds of cuts through
the lattice \cite{Bonnet:2000kw}.) As can be seen from Fig.~\ref{Latt} an 
O(4) invariant cutoff of the sums necessitates to neglect the ``edges'':
The sum extends only over the fully drawn hypersheres, and we omit the summation
over the dashed ones.

To be able to compare to already known flat space solutions we will first solve
the torus analogue of the equations with angle approximations, {\it i.e.} eqs.\
(\ref{odGDSE},\ref{odZDSE}) or eqs.\ (\ref{ghAB},\ref{glAB}). To this end we
start from eqs.\ (\ref{ghDSE1},\ref{ZDSE}) or eqs.\ (\ref{ghost},\ref{gluon1}),
respectively. After formulating the corresponding four-dimensional integrals
as sums over lattice momenta the functional dependence of the propagator 
functions is approximated accordingly to the rules described in the subsections
II.A and II.B, respectively.
We give the explicit expressions for the bare vertex ghost-loop only
truncation. The corresponding ones for dressed vertex truncation can be derived
analogously in a straightforward manner, however, as these are quite lengthy we
do not give their explicit form.

The Dyson--Schwinger equations in bare vertex ghost-loop only truncation with 
angle approximation read on the torus
\begin{eqnarray}
\frac{1}{G(x_i)} &=&  \widetilde{Z}_3(s,L) - {g^2}{N_c} \frac{1}{L^4} \sum_j 
\frac{K(x_i,y_j,z_n)}{x_i y_j} G(y_j) Z({\rm max}(x_i,y_j))
 \; ,
\label{disGDSE}
\\
\frac{1}{Z(x_i)}&=& Z_3 (s,L) + {g^2} \frac{N_c}{3} 
\frac{1}{L^4} \sum _j \frac{M(x_i,y_j,z_n)}{x_i y_j} G(y_j) 
G({\rm max}(x_i,y_j))   \; . 
\label{disZDSE} 
\end{eqnarray} 
As shown in the previous subsection the momentum $p=k-q$ which fulfills $z_n=
(k-q)^2$ lies on the momentum grid, and furthermore,  $p=k-q$ is determined
uniquely from $k$ and $q$ for which $x_i=k^2$ and $y_j=q^2$ by using elementary
vector operations. Note, however, that $\sqrt{p^2}$ might be larger than the
ultraviolet cutoff even if $\sqrt{k^2}$ and $\sqrt{q^2}$ are not. The actual
treatment of the sums in eqs.\ (\ref{disGDSE},\ref{disZDSE}) can nevertheless
be done straightforwardly as the exact algebraic expressions for the kernels
are known.

The main difference between the equations on the torus and the flat space
equations is the effective treatment in the infrared. The finite volume in
coordinate space leads to a finite value of squared momentum for the first
hypersphere $j=1$. Thus one has not to worry about possible infrared
singularities. On the other hand, there exists zero modes on the four-torus. In
all the calculations presented here they are neglected, an estimate of
their possible contribution is given in Appendix B. Note furthermore that for
the numerical solution as described in Refs.\
\cite{vonSmekal:1997is,Hauck:1998sm} the disadvantage of dealing with infrared
singularities has been turned  into a benefit: The infrared behavior of the
propagator functions has been determined via an asymptotice series calculating
the corresponding coefficients and powers analytically. Matching these series
at some finite momentum to the functions  as obtained from standard numerical
techniques finally enabled to numerically solve the coupled integral equations.
Such a complicated process is not necessary employing a torus. 
However, one  anticipates already at this level some
deviations in the infrared between the solutions obtained in these different
ways. In the next section the corresponding numerical results will be dicussed.
They demonstrate that using a torus as infrared cutoff works surprisingly well.

\subsection{B. Dyson--Schwinger equations with angle integrals on the torus}

The intricacy of the infrared analysis has prevented so far a solution of the
coupled gluon-ghost system without angular approximations. 
Using a torus as an infrared regulator opens up the 
possiblity to solve these equations directly without any recourse to an 
infrared asymptotic expansion. As recently arguments have been provided that
the use of a bare ghost-gluon vertex is not inferior to employing a dressed one
\cite{Zwanziger:2001kw,Lerche:2001} we will solve the system in the truncation
described in sect.~II.C.

With the above discussed replacement of $\int \frac{d^4q}{(2\pi)^4} \rightarrow
\frac{1}{L^4} \sum_j$ the equations (\ref{ghostbare},\ref{gluonbare}) read
on the torus:
\begin{eqnarray}
\frac{1}{G(x_i)} &=& \widetilde{Z}_3(s,L) - {g^2}{N_c}\frac{1}{L^4} \sum_j 
\frac{K(x_i,y_j,z_n)}{x_i y_j} G(y_j) Z(z_n) \; ,
\label{ghost_tor} \\
\frac{1}{Z(x_i)} &=& {Z}_3 (s,L) + {g^2} \frac{N_c}{3}
\frac{1}{L^4} \sum _j \frac{M(x_i,y_j,z_n)}{x_i y_j} G(y_j) G(z_n)
\label{gluon_tor}\\
&& \:\:\:\:\:\:\:\: + {g^2} \frac{N_c}{3}
\frac{1}{L^4} \sum_j \frac{Q^\prime(x_i,y_j,z_n)}{x_i y_j} 
G^{1-a/\delta-2a}(y_j) G^{1-b/\delta-2b}(z_n)
Z^{-a}(y_j) Z^{-b}(z_n) \; .\nonumber
\end{eqnarray}
As already stated $\sqrt{z}$ might be larger than the
ultraviolet cutoff even if $\sqrt{x}$ and $\sqrt{y}$ are not.
Again the kernels can be nevertheless evaluated straightforwardly, however,
if $z$ with $L<z<4L$ is the argument of a propagator function
we set $z=L$, {\it i.e.\/} we approximate $Z(z)$ or $G(z)$ then by
$Z(L)$ or $G(L)$, respectively. Another more elaborate treatment consists of 
matching the corresponding perturbative ultraviolet tail to the function under 
consideration. This has been applied in some cases to test the viability of the 
method.

\goodbreak

\section{VI. Numerical results}

\subsection{A. Comparing torus solutions to previous results in flat space-time}

\begin{figure}
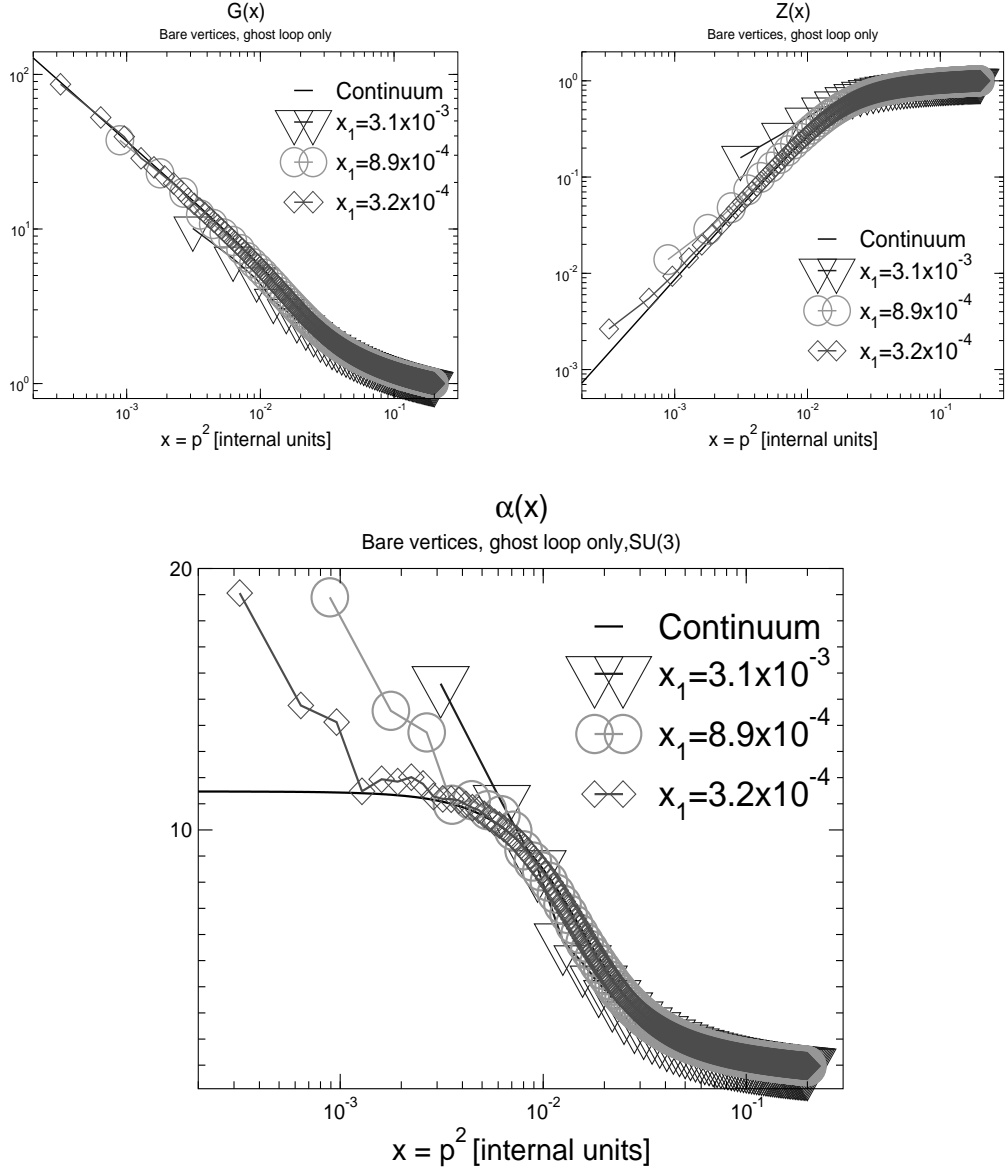

\centerline{
\epsfig{file=bare_ghostonly.g.eps,width=6.0cm,height=6cm}
\hspace{1cm}
\epsfig{file=bare_ghostonly.z.eps,width=6.0cm,height=6cm}
}

\vspace{0.5cm}
\centerline{
\epsfig{file=bare_ghostonly.a.eps,width=9cm,height=9cm}
}

\caption{\label{bare.dat}
Shown are the ghost dressing function, the gluon dressing function and
the running coupling in the bare-vertex ghost-only truncation for
different momentum grid spacings corresponding to different 
finite volumes of the torus. The fully drawn lines labeled ``continuum''
represent the respective results for flat Euclidean space-time, {\it i.e.\/}
continuous momenta.}
\end{figure}

The solutions of the Dyson--Schwinger equations on the torus can be compared in
two different ways to the ones obtained in flat space-time. First, one can use
the solutions of the ``continuum'' equations for a certain cutoff $\Lambda$ and
certain renormalization scale $\mu$ to provide the  values for
$Z_3(\mu^2,\Lambda^2)$ and $\tilde{Z}_3(\mu^2,\Lambda^2)$ as input for the
equations on the torus which are then solved, {\it c.f.\/} the discussion in
Appendix C where the numerical methods employed to obtain the solutions of the
``continuum'' equations are summarized.  Second, one can subtract both equations
at the squared momenta $s_G$ and $s_Z$ and therefore trade the two
renormalization constants for the values of the dressing functions at these
momenta, namely $Z(s_Z)$ and $G(s_G)$. For the sake of comparison one can read
of these values from the continuum solution to be used in the torus equations.
If $s_Z$ and $s_G$ are taken to be sufficently far in the ultraviolet region of
momentum, where finite volume effects play a minor role, the two procedures
lead to the same results. We have verified that this is indeed the case within
the limits of numerical accuracy.

\begin{figure}
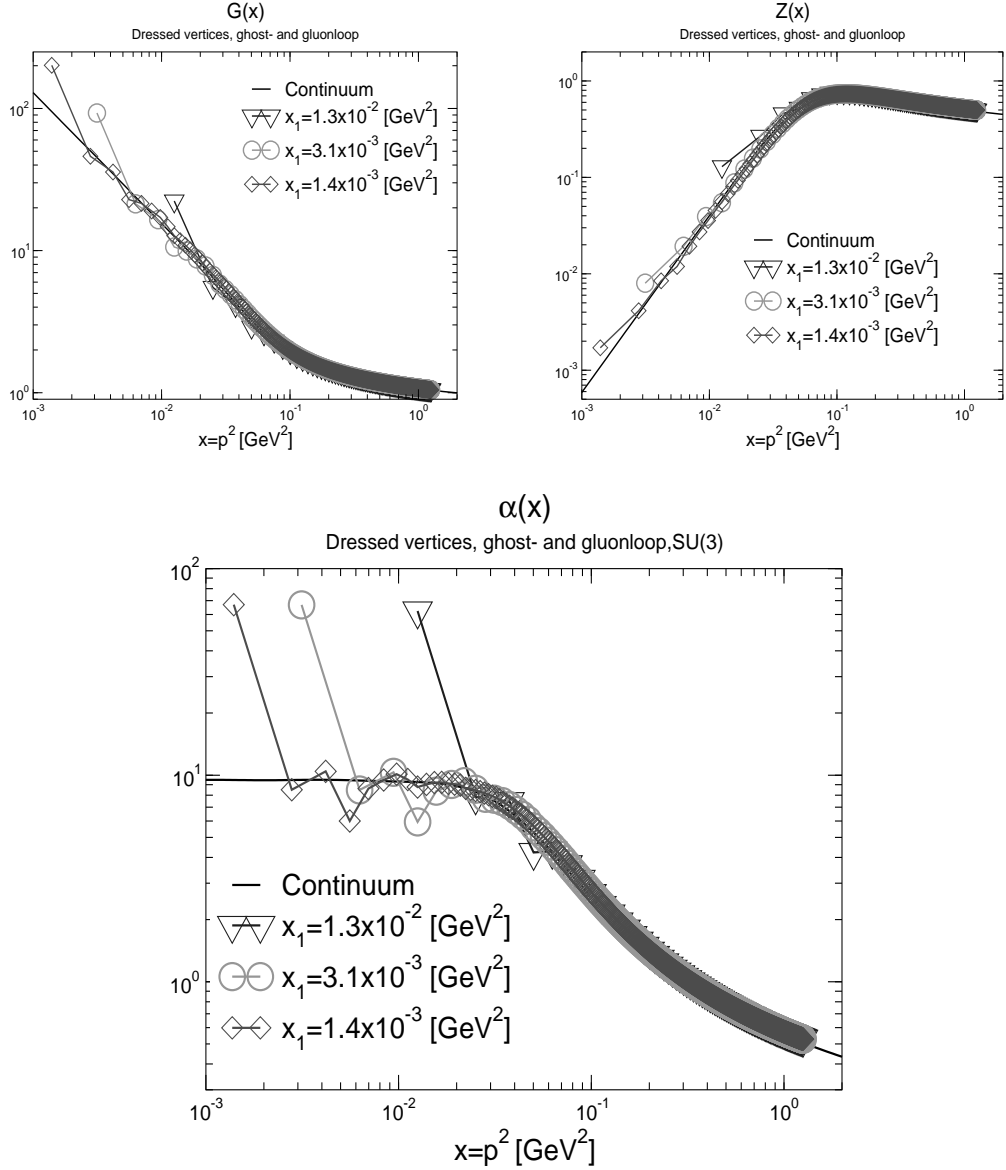

\centerline{
\epsfig{file=dressed_gluonincl.g.eps,width=6.0cm,height=6cm}
\hspace{1cm}
\epsfig{file=dressed_gluonincl.z.eps,width=6.0cm,height=6cm}
}

\vspace{0.5cm}
\centerline{
\epsfig{file=dressed_gluonincl.a.eps,width=9.0cm,height=9cm}
}

\caption{\label{dressed.dat}
The same as fig.\ \protect{\ref{bare.dat}} for the dressed vertex truncation.}
\end{figure}

Our results for the ghost dressing function, the gluon dressing function and
the running coupling in the bare-vertex ghost-only and the dressed vertex
truncation can be seen in Figs.\ \ref{bare.dat} and \ref{dressed.dat},
respectively. In both truncation schemes we solved for three
different momentum spacings corresponding to different volumes in coordinate
space. To keep the cutoff identical for all the spacings within each truncation
scheme we have chosen three different grid sizes respectively. For the bare
vertex truncation they are $N=17^4, 31^4, 51^4$ and for the dressed vertex
truncation they are $N=21^4, 41^4, 61^4$.

A physical momentum scale cannot be determined in truncation schemes which do
not provide the correct perturbative running of the coupling in the ultraviolet.
With the missing gluon loop this is the case in the bare vertex
only truncation scheme. We therefore have to stick to an internal momentum
scale without physical units in this case. The situation is different, however,
in the dressed vertex truncation scheme. Here we have fixed the momentum scale
by calculating the running coupling for the colour group SU(3) and using the
experimental value $\alpha(x)=0.118$ at $x=M_Z^2=(91.187 \mbox{GeV})^2$ 
to fix a physical scale.

In the bare vertex truncation scheme we have chosen $Z(\mu^2)=G(\mu^2)=1$ as
renormalization condition where we have taken $\mu^2=\Lambda^2$, the (squared)
ultraviolet cutoff. Of course, this choice is by no means special and one is
completely free to choose the renormalization point wherever one likes. For the
numerical calculation in flat space we have chosen two different subtraction
points for the ghost and the gluon equations as described in Appendix C. For
good convergence of the iteration process the ghost equation is most
conveniently subtracted at zero momentum, whereas the gluon equation can be
subtracted at any value of squared momentum in the region where the equations
are solved numerically. In our calculation we chose the cutoff $\Lambda^2=0.2$
in internal units as subtraction point for the gluon equation.  This allows us
to use the renormalization condition $Z(\Lambda^2)=1$ directly as input in the
calculation. The second input is provided by the coefficient $A$ of the leading
order infrared expansion of the gluon dressing function, $Z_{IR}(x)=Ax^{2
\kappa}$. The condition $G(\Lambda^2)=1$ then leads to $A=357.33$. The value of
the coupling at the renormalization point, $\alpha(\mu^2)$, is taken to be
0.97. Again one is completely free to chose this number arbitrarily up to the
maximum value of the running coupling which is reached in the very infrared. 
For the torus calculations the unsubtracted equations have been used. The same 
coupling $\alpha(\mu^2)=0.97$ has been taken and instead of $Z(\Lambda^2)$ and
$A$ the values $Z_3(\mu^2=0.2,\Lambda^2=0.2)=0.9591$ and
$\tilde{Z}_3(\mu^2=0.2,\Lambda^2=0.2)=1.1034$ for the renormalization constants
determined from the ``continuum'' solution have been employed.

As has already been mentioned, in the dressed vertex truncation scheme we
determine a ``physical'' momentum scale according to experimental results. In
the numerical treatment again the ghost equation has been subtracted at zero
momentum whereas the gluon equation is subtracted at finite momentum. We solved
both equations similar to the method described in Ref.\ \cite{Hauck:1998sm},
especially we introduced also the auxiliary functions $F(x)$ and $R(x)$ as
defined in Ref.\ \cite{vonSmekal:1997is}. 
As input values serve the infrared expansion of $R(x)$, $R(x)=x^\kappa +\ldots$,
and the value $R(s)=0.8$ at the gluon subtraction point
$s=1.048$GeV$^2$. For the calculations on the torus we use the unsubtracted
equations with the values $Z_3(\mu^2=M_Z^2,\Lambda^2=1.255
\mbox{GeV}^2)=1.266$ and $\tilde{Z}_3(\mu^2=M_Z^2,\Lambda^2=1.255
\mbox{GeV}^2)=0.966$ for the renormalization constants which have been
determined from the ``continuum'' solution.

Our numerical results, depicted in Figs.\ \ref{bare.dat} and \ref{dressed.dat},
show similar properties for both truncation schemes. Compared with the
respective ``continuum'' solutions the ones obtained on a torus show deviations
for the first few lattice points in the infrared. For higher momenta all
functions on the torus approach the continuum ones. Deviations are there only
visible for the curves with the largest spacings. The biggest effect can be
seen for the running coupling $\alpha$, which is the ``observable'' of the
system. As $\alpha$ is proportional to the product $Z(x)G^2(x)$ the deviations of
the dressing functions from the ``continuum'' curve amplify in the infrared in
a somewhat erratic way, so that the points in the very infrared cannot be
connected by a smooth line. Comparing larger and smaller spacings of momentum
grids one clearly sees that the effect is always one of the first spheres on
the respective lattices and therefore moves to the infrared for smaller
spacings. 
%The continuum functions are sort of 'envelopes' for the torus
%functions with different spacings.

The most important properties of the solutions in flat space can still be 
found in the torus solutions despite some deviations in the infrared. Going
from larger to smaller spacings a powerlike behaviour of the dressing functions
in the infrared with the correct exponents can still be infered. For the truncation
scheme with dressed vertices the gluon dressing function on the torus has the
same shape and the same hight of the bump in the bending region of the curve.
As all curves with sufficently small momentum spacing follow the correct power
behaviour within numerical errors we conclude that one surely can adress the
question of the value of the power $\kappa$ in the infrared on a manifold with
finite volume as has been used here or as one uses in lattice calculations.
This is the central result of the present section: Employing a torus as
infrared regularization is possible. In the following section we will use
this to go beyond the angular approximation.

\subsection{B. Numerical solutions without angular approximations}

In this section we discuss our numerical results for the bare vertex truncation
scheme with various projectors in the gluon equation. We show results for the
gluon and ghost dressing function and the running coupling calculated on a
torus without any angular approximations. Note that up to now in the
gluon-ghost Dyson--Schwinger equations solutions for non-vanishing momenta have
been obtained only employing some sort of angular approximation. Going beyond
these approximations we will use our new truncation scheme introduced in sect.\
2.3. The reason for this is, that the dressed vertex truncation scheme is 
quite complicated, and in its
derivation one has dismissed terms which seem to be crucial for the infrared
behavior beyond the angle approximation. As the survey of gluon and ghosts 
infrared behaviors related to different ans\"atze for vertex functions
\cite{Lerche:2001} does not provide any evidence for crucial differences
between bare and dressed vertex functions the use of the latter seems to be an
unnecessary complication. 
In the case of the bare-vertex ghost-loop only truncation with
Brown--Pennington projector no solutions for finite momenta without angular
approximation have been reported yet. In fact the results presented in this 
section suggest that these solutions might not exist.

This time the values for the renormalization constants $Z_3$ and $\tilde{Z}_3$
cannot be taken from a ``continuum'' solution. Thus we use the subtracted 
equations on the torus in the course of the numerical solution by iteration.
Both equations are subtracted at the renormalization point $\mu^2=1.9
\mbox{GeV}^2$. Similar to the case of the dressed vertex truncation from the
last section the 'physical' units are gained from experimental input:
$\alpha(x)=0.118$ at $x=M_Z^2=(91.187 \mbox{GeV})^2$. The input values for the 
dressing functions at the renormalization point are $Z(\mu^2)=0.83$. Requiring
$G(\mu^2)=1/\sqrt{Z(\mu^2)}$ then fixes the overall scale for $G(x)$. For the
value of the coupling at the renormalization point we chose again
$\alpha(\mu^2)=0.97$ similar to the two truncation schemes in the last section.
For the presented solution on three different volumes lattice sizes of $N=13^4,
43^4,71^4$ have been used.

\begin{figure}
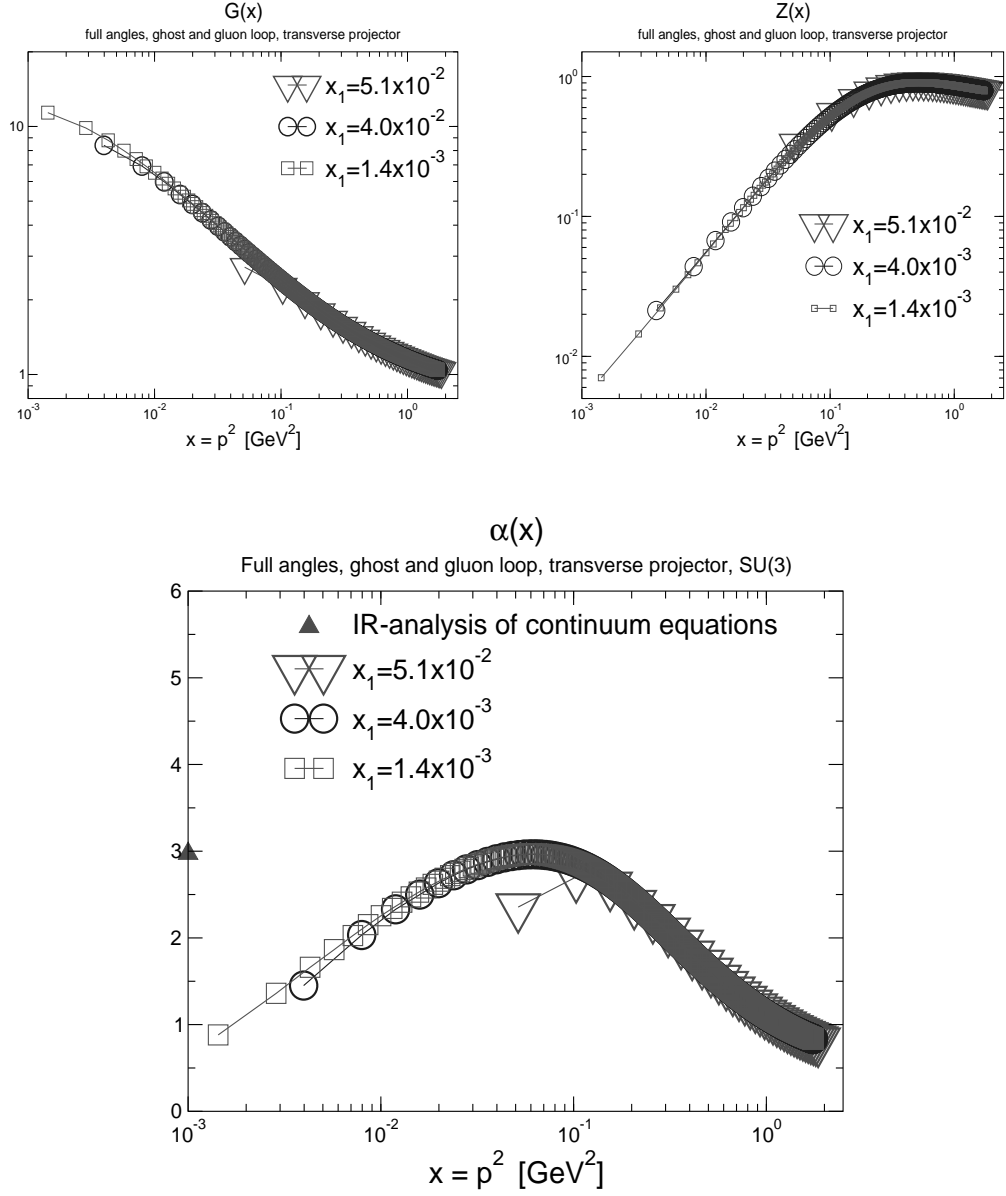

\centerline{
\epsfig{file=newtrunc.main.g.eps,width=6.0cm,height=6cm}
\hspace{1cm}
\epsfig{file=newtrunc.main.z.eps,width=6.0cm,height=6cm}
}

\vspace{0.8cm}
\centerline{
\epsfig{file=newtrunc.main.a.eps,width=9.0cm,height=9cm}
}

\caption{\label{new.dat}
Shown are the ghost dressing function, the gluon dressing function and
the running coupling in the new truncation scheme without angular 
approximation for different volumes using a transverse projector.}
\end{figure}

\begin{figure}
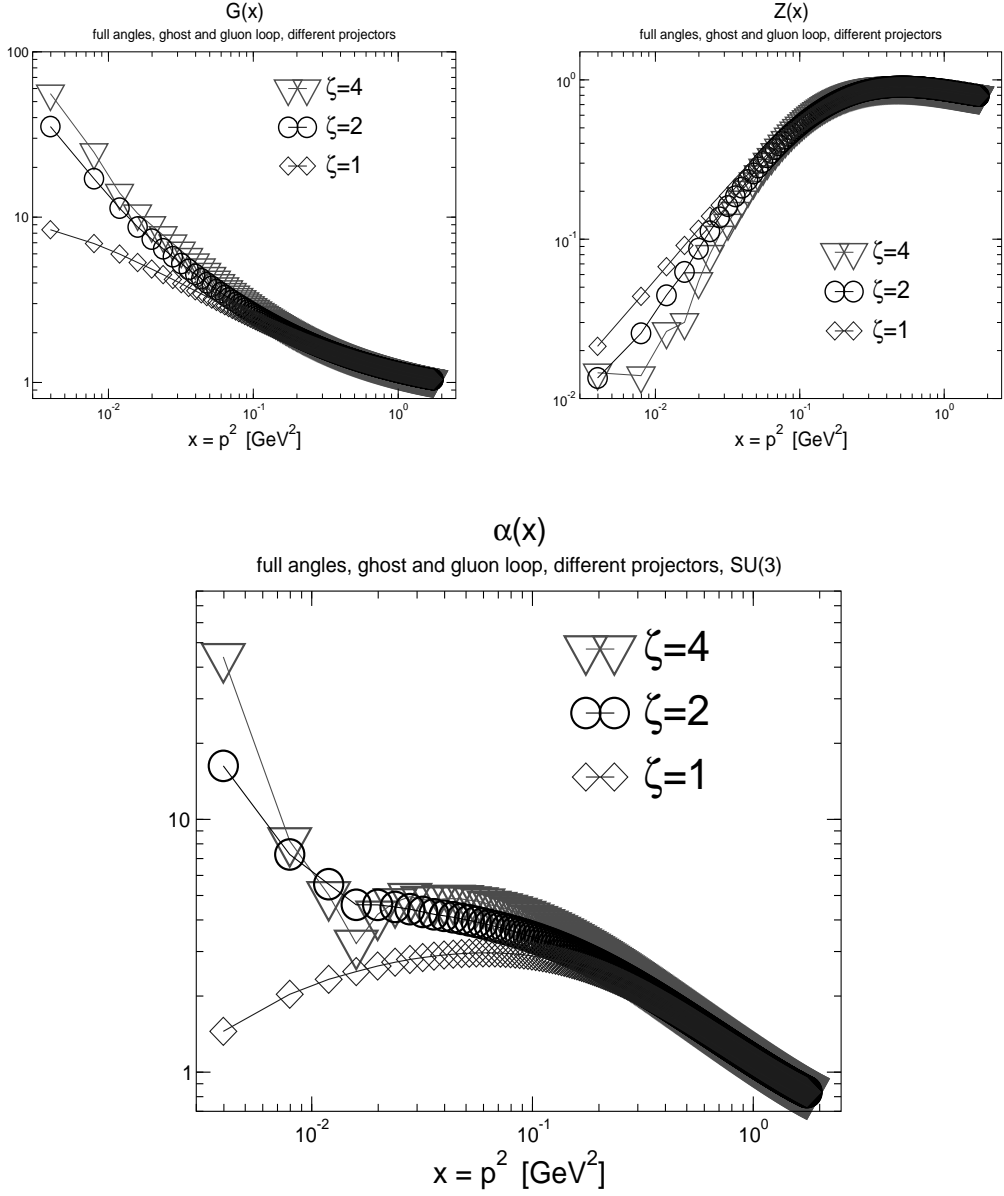

\centerline{
\epsfig{file=newtrunc.proj.g.eps,width=6.0cm,height=6cm}
\hspace{1cm}
\epsfig{file=newtrunc.proj.z.eps,width=6.0cm,height=6cm}
}

\vspace{0.8cm}
\centerline{
\epsfig{file=newtrunc.proj.a.eps,width=9.0cm,height=9cm}
}
\caption{\label{proj.dat}
The same as fig.\ \protect{\ref{new.dat}} for different projectors.}
\end{figure}

As mentioned in the last section we did check cutoff effects by extrapolating
the propagator functions at $z > \Lambda^2$ with a logarithmic tail with the 
correct anomalous dimensions. The results as compared to the ones obtained by 
simply setting $Z(z)=Z(\Lambda^2)$ and $G(z)=G(\Lambda^2)$ for  
all $z>\Lambda^2$ did change by less than $10^{-3}$.

Our results for the dressing functions obtained with a transverse projector as
shown in Fig.\ \ref{new.dat} are in agreement with the expected power behavior.
The infrared critical exponent as calculated in Refs.\ 
\cite{Zwanziger:2001kw,Lerche:2001}, $\kappa=0.5953$, thus has been verified: A
corresponding (numerical) solution for non-vanishing momenta exists. The gluon
dressing function is remarkably stable against changes of the volume  and
approaches more and more the expected power solution for small momenta.  For
the ghost dressing function one observes again deviations of the first points
in the infrared: An extraction of the correct infrared critical exponent from
the numerical solution for the ghost function is hardly possible. Only some
points come close to the analytical value of the continuum before the curve
starts bending down again in the very infrared. For the extracted value of the
running coupling in the infrared this leads to a distinct mismatch to what one
is to expect on the basis of analytical results.

At first sight the fact that the power solution for the ghost dressing function
could be not reproduced numerically to a reasonable precision may seem
disappointing. Nevertheless these numerical results themselves show that the
ghost dressing function is highly infrared singular. This reflects the
long-range correlation of ghosts in Landau gauge. Therefore one should expect
the ghost dressing  function  to be the one  affected most by a finite volume.
On the contrary the gluon dressing function vanishes in the infrared and
consequently it is much less affected by a finite volume. We expect the ghost
dressing function together with the running coupling to approach more and more
the correct power solution in the infrared as lattice spacings are decreased
and lattice sizes are increased.

Furthermore, we add a remark on the transversality of the gluon propagator as
obtained in the bare vertex truncation scheme. Of course, the full gluon
propagator calculated from the complete gluon equation  with fully dressed
vertices in Landau gauge is transversal due to Slavnov--Taylor identities. 
Thus such a hypothetical solution would be independent of the form of the
projector, {\it i.e.\/} in our notation with the projector 
$(\delta_{\mu \nu} - \zeta p_\mu p_\nu)/p^2$ independent of the parameter $\zeta$. In practice, using
bare vertices as in our truncation scheme this is certainly not the case. Our
numerical results for different values of the parameter $\zeta$ can be seen in
Fig.\ \ref{proj.dat}. Although our solutions show the expected dependence
on the form of the projector this dependence is not too drastic and in general
the behaviour of these different solutions is very similar.

For the gluon dressing function one observes that the more $\zeta$ grows the
greater is the deviation from the pure power behaviour. The points in the very
infrared cannot be connected by a smooth line any more. Correspondingly this
happens for the running coupling. Based on the infrared analysis  one might
anticipate
that $\kappa$ should approach the value $\kappa=0.5$ more and more as $\zeta$
grows until there is a jump to the solution from $\kappa=0.5^+$ to $\kappa=1$
as the Brown--Pennington limit $\zeta=4$ is reached. We do not observe such
a qualitative jump in our solutions on a torus. The  solution shown for
$\zeta=4$  is approached smoothly when $\zeta$ approaches this limit. 
This clearly indicates that the solution $\kappa=1$ might not exist at all
if one removes the torus as a regulator.

\begin{figure}
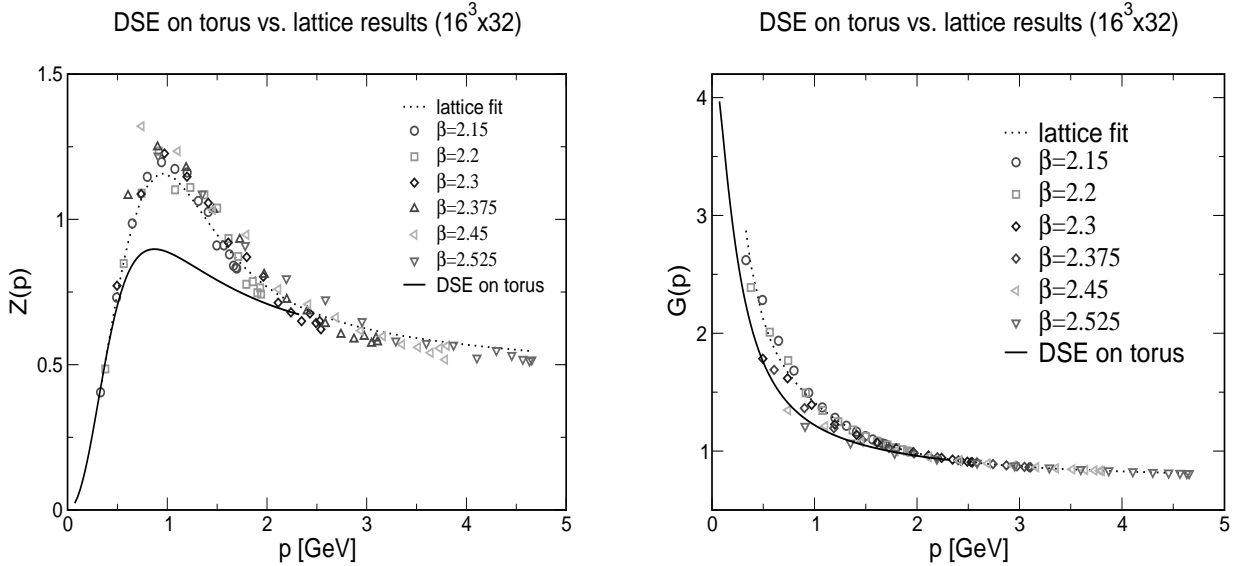

\centerline{
\epsfig{file=latcomp.glue.eps,width=7.5cm,height=7.5cm}
\hspace{1.0cm}
\epsfig{file=latcomp.ghost.eps,width=7.5cm,height=7.5cm}
}
\caption{\label{lattice.dat}
Results on the torus compared to recent lattice results \cite{Langfeld:2002}. 
As the torus points are very
close to each other on a linear momentum scale we did not resolve the torus curves into
single points.}
\end{figure}

Finally we compare our results to recent SU(2)
lattice calculations \cite{Langfeld:2002}. As has already been stated above 
the ghost and gluon dressing functions from Dyson-Schwinger equations are independent
from the numbers of colors at least to our level of truncation. The only caveat in
comparing our results with the lattice ones is the adjustment of the momentum
scale which is certainly different from the case of SU(3) above. To get
an appropriate momentum scale we therefore
used the lattice result $\alpha_{SU(2)}(x)=0.68$ at $x=10 \mbox{GeV}^2$ as input. 
The two graphs in figure (\ref{lattice.dat}) show that the main
qualitative features, the vanishing of the gluon and the divergence of
the ghost dressing functions in the infrared, are common properties of both, the
lattice solutions and the one from Dyson-Schwinger equations. Even the power 
$\kappa=0.59...$ of the gluon dressing function on the torus is very close 
to the one that can be extracted from the lattice fit to be $\kappa \approx 0.5$.
The main difference of both approaches lies in the medium energy region around one
GeV, where the Dyson-Schwinger solutions suffer from the missing two loop contributions
that are certainly present in lattice Monte-Carlo simulations.

\section{V. Conclusions and Outlook}

In this paper we have presented numerical solutions of truncated systems of
Dyson--Schwinger equations for the gluon and ghost propagators in Landau gauge
SU(N) Yang--Mills theories. We have employed a four-torus, {\it i.e.}~a
compact space-time manifold, as an infrared regulator. This enabled us to
overcome angular approximations used so far in previous studies
\cite{vonSmekal:1997is,Atkinson:1998tu}. The basis for our numerical
calculations have been provided by an analytic determination
\cite{Zwanziger:2001kw,Lerche:2001} of the exponents governing the infrared
power-like behavior of the gluon and ghost propagators. Typically  the infrared
analysis provide two possible values each for the gluon and the  ghost,
respectively. The central result of this paper is: Only one of these  two
possible infrared behaviors in a given truncation scheme can be matched to a
numerical solution for non-vanishing momenta. 

In a truncation scheme with a bare  ghost-gluon vertex and a transverse
projection of the (truncated) gluon equation this implies that the gluon
propagator is only weakly infrared vanishing, $D_{gl}(k^2) \propto
(k^2)^{2\kappa -1}$, $\kappa =0.59\ldots$, and the ghost propagator is 
highly infrared
singular, $D_{gh}(k^2) \propto (k^2)^{-\kappa -1}$. The running coupling
possesses an infrared fixed point whose value is given by $\alpha(0) \approx
2.97$ (or, for a general number $N_c$  of colors, $\alpha(0) \approx 8.92/N_c$). 

Our results compare very favorably with results of recent lattice 
calculations performed
for two colors. Due to the finite lattice volume the lattice results cannot, of
course, be extended into the far infrared. In this respect our results are
complementary to the lattice ones: We do obtain the infrared behavior
analytically. On the other hand, lattice calculations include, at least in
principle, all non-perturbative effects whereas we had to rely on truncations.
{\it E.g.\/} the deviations for the gluon renormalization functions at
intermediate momenta depicted in Fig.\ \ref{lattice.dat}
might be due to the neglect of the four-gluon vertex
function in our calculations.

As an outlook we would like to mention that employing a four-torus as underlying
mani\-fold might serve for a number of interesting studies of Dyson--Schwinger
equations. Amongst these, the use of 
twisted boundary conditions on the torus might enable one to shed some
light on the importance of topologically non-trivial gauge field configurations
for the infrared behavior of QCD Green's functions. We would like to note that
recent lattice calculations \cite{Langfeld:2001cz} indicate a relation between
the existence of center vortices in the maximal center gauge, the area law of
the Wilson loop and the infrared behavior of the gluon propagator in Landau
gauge.

Furthermore, we want to point out that choosing an asymmetric four-torus 
might provide an efficient mean to extend these calculations to non-vanishing
temperatures. Here the main qualitative question arises about the fate of the
Kugo--Ojima confinement criterion at the deconfinement transition.

Finally, we want to remark that QCD Green's functions are an important input in
many calculations in hadron physics \cite{Alkofer:2001wg,Roberts:2000aa}.
The next necessary step towards such phenemenological applications is a study of
the quark propagator. Such an investigation might hopefully also provide some
insight into the mechanism of quark confinement in covariant gauges. 

\section*{Acknowledgements}

We thank Jacques Bloch for many useful hints in the early stages of this work.
We are especially grateful to Lorenz von Smekal for many valuable discussions and 
for making ref.\ \cite{Lerche:2001} available to us. We are indebted to Kurt Langfeld
for communicating and elucidating his lattice results, partly prior to publication.

Furthermore, we thank Sebastian Schmidt, 
Peter Watson, Andreas Wipf and Daniel Zwanziger for helpful discussions. 

This work has been supported by the DFG under contract Al 279/3-3 and by the
European graduate school T\"ubingen--Basel.

\newpage

\begin{appendix}

\section{Appendix A: One-loop scaling}

In the framework of the truncation scheme presented in sect.~IIC
we have shown that the approximation
\begin{equation} 
{\cal Z}_1(x,y,z;s,L) = 
\frac{G(y)^{(1-a/\delta-2a)}}{Z(y)^{(1+a)}}
\frac{G(z)^{(1-b/\delta-2b)}}{Z(z)^{(1+b)}}
\end{equation}
for the gluon vertex renormalization constant yields the correct one loop
scaling of the gluon loop in the gluon Dyson-Schwinger equation. This is
true for any values $a$ and $b$. Of course, in a full treatment of the
coupled ghost gluon system ${\cal Z}_1(s,L)$ would be independent
of momentum. Therefore a choice of $a$ and $b$ which keeps ${\cal Z}_1$ as
weakly varying as possible seems the most reasonable one.
\begin{figure}
%\vspace{-2cm}
\begin{center}
\epsfig{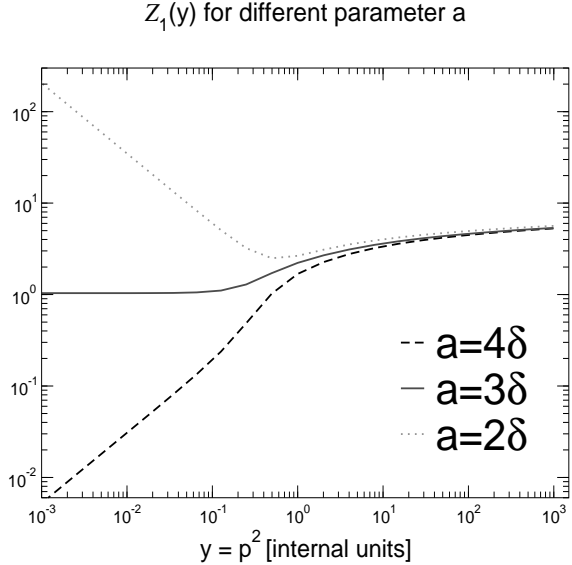}
\end{center}
\caption{\label{z1.dat}
The $y$-dependence of the function ${\cal Z}_1(y,z)$ for different
values of the parameter $a$. Only the choice $a=3\delta$ leads to 
momentum independence in the infrared. Due to the symmetry of the ansatz for
${\cal Z}_1(y,z)$ the $z$-dependence is the same for $b=a$.
}
\end{figure}

This choice can be infered using the scaling of the dressing functions
extracted from the renormalization group equation, see ref.~\cite{vonSmekal:1997is}
for details. The dressing functions can be expressed as
\begin{eqnarray}
Z(x) &=& \left(\frac{\alpha(x)}{\alpha(s)}\right)^{1+2\delta}R^2(x) \; ,
 \nonumber\\
G(x) &=& \left(\frac{\alpha(x)}{\alpha(s)}\right)^{-\delta}R^{-1}(x) \; ,
\end{eqnarray}
where the running coupling provides the correct one loop scaling in the ultraviolet.
Consequently the function $R(x)$ approaches unity for high momenta.
Furthermore, from the known infrared behavior of $Z(x)$, $G(x)$ and
$\alpha(x)$ one infers that $R(x)$ 
is proportional to $x^\kappa$ in the infrared. Writing ${\cal Z}_1$ in terms of $\alpha(x)$
and $R(x)$ yields
\begin{equation} 
{\cal Z}_1(x,y,z;s,L) = 
\left(\frac{\alpha(\mu^2)}{\alpha(y)}\right)^{1+3\delta} R^{-3+a/\delta}(y)
\left(\frac{\alpha(\mu^2)}{\alpha(z)}\right)^{1+3\delta} R^{-3+b/\delta}(z).
\end{equation}
In the perturbative region $R(y),R(z) \rightarrow 1$ and the function ${\cal Z}_1$ is
therefore slowly varying for any $a$ and $b$ 
according to the logarithmic behaviour of the running coupling 
$\alpha$. In the infrared, however, $\alpha$ approaches its fixed point while
the functions $R$ behaves like a power. Consequently the choice $a=b=3\delta$ 
guarantees the weakest momentum dependence of ${\cal Z}_1$ which is illustrated
in Fig.~\ref{z1.dat}.
Shown is the $y$-dependence of the function
${\cal Z}_1(x,y,z=y;s,L)$. (Note also that due the symmetry 
${\cal Z}_1(x,y,z;s,L)={\cal Z}_1(x,z,y;s,L)$ and the absence of an explicit
$x$-dependence this is sufficient to demonstrate
its momentum dependence.)
In the perturbative momentum regime the function ${\cal Z}_1$ does not vary
with the parameter $a$. So all three choices give the same logarithmic running
in momentum as required to give the correct one loop scaling behaviour of the
integral. In the infrared, however, a change in  $a$ gives rise to substantial
changes in the behaviour of ${\cal Z}_1$, with only the choice $a=3\delta$
leading to a constant.

\section{Appendix B: The influence of zero modes on the solutions}

An important point when formulating the Dyson--Schwinger equations  on the
torus could be the treatment of the zero modes.  In addition, on the torus an
infrared analysis like the one in flat Euclidean space-time is not possible,
and one is left with the problem how the dressing functions behave at 
vanishing momenta. Guided by the intuition that especially the
long ranged modes should be affected by the  finite volume we assume in the
following $Z(x \rightarrow 0) = 0$ just like in the continuum and $G(x
\rightarrow 0) =const$ if zero modes are neglected. Phrased otherwise we
assume that the zero modes are the missing ingredient to ensure the correct
infinite volume limit for the torus results. Therefore, if on tori of different
volumes $G(x= 0)$ shows no sign of becoming divergent, the infrared enhancement
seen in $G(x \rightarrow 0)$ or in the flat space-time results has to be due to
the torus zero modes of gluons and ghosts.

Therefore, in this appendix, we will show that the  assumption $G(x= 0) <
\infty$ does not lead to a contradiction in the equations on the torus if zero
modes are neglected. To this end we focus on the truncation scheme without
angular approximations.  First we rewrite equations
(\ref{ghostbare},\ref{gluonbare}) as
\begin{eqnarray} 
\frac{1}{G(x)} &=& Z_3 - g^2N_c \int \frac{d^4q}{(2 \pi)^4}
\frac{K(x,y,z)}{xy}
G(y) Z(z) \; , \label{ghostbare_app} \\ 
\frac{1}{Z(x)} &=& \tilde{Z}_3 + g^2\frac{N_c}{3} 
\int \frac{d^4q}{(2 \pi)^4} \frac{M(x,y,z)}{xy} G(y) G(z) \nonumber\\
&&+ g^2 \frac{N_c}{3} \int \frac{d^4q}{(2 \pi)^4} 
\frac{Q(x,y,z)}{xy} \frac{G(y)^{-2-6\delta}}{Z(y)^{3\delta}} 
\frac{G(z)^{-2-6\delta}}{Z(z)^{3\delta}} \; .\nonumber\\
\label{gluonbare_app} 
\end{eqnarray} 
According to Appendix A we have chosen $a=b=3\delta$, where $\delta=-9/44$, the
anomalous dimension of the ghost. The kernels have the form:
\begin{eqnarray}
K(x,y,z) &=& \frac{1}{z^2}\left(-\frac{(x-y)^2}{4}\right) + 
\frac{1}{z}\left(\frac{x+y}{2}\right)-\frac{1}{4} = xy\frac{\sin^2\Theta}{z^2}
\; , \\
M(x,y,z) &=& \frac{1}{z} \left( \frac{\zeta-2}{4}x + 
\frac{y}{2} - \frac{\zeta}{4}\frac{y^2}{x}\right)
+\frac{1}{2} + \frac{\zeta}{2}\frac{y}{x} - \frac{\zeta}{4}\frac{z}{x}\; ,  \\
Q^\prime(x,y,z) &=& \frac{1}{z^2} 
\left( \frac{1}{8}\frac{x^3}{y} + x^2 -\frac{19-\zeta}{8}xy + 
\frac{5-\zeta}{4}y^2
+\frac{\zeta}{8}\frac{y^3}{x} \right)\nonumber\\
&& +\frac{1}{z} \left( \frac{x^2}{y} - \frac{15+\zeta}{4}x-
\frac{17-\zeta}{4}y+\zeta\frac{y^2}{x}\right)\nonumber\\
&& - \left( \frac{19-\zeta}{8}\frac{x}{y}-\frac{3-4\zeta}{2}+
\frac{9\zeta}{4}\frac{y}{x} \right) \nonumber\\
&& + z\left(\frac{\zeta}{x}+\frac{5-\zeta}{4y}\right) + z^2\frac{\zeta}{8xy}
\; .
\label{new_kernels_app}
\end{eqnarray}

We first analyse the behavior
of the integrands in the limit $y \rightarrow 0$ for finite momenta $x$. 
Then $Z(z) \rightarrow Z(x)$ and $G(z) \rightarrow G(x)$ and 
the kernels times the respective dressing functions are 
to appropriate order in momentum $y$:
\begin{eqnarray}
\frac{G(y) Z(z)}{xy}K(x,y,z) &\rightarrow& G(0)Z(x)\frac{\sin^2\Theta}{x^2}\; , 
\label{kernel_y1}\\
\frac{G(y) G(z)}{xy}M(x,y,z) &\rightarrow& G(0)G(x) \frac{1}{x^2} 
\left(1+(\zeta-2)\cos^2\Theta \right) \; , 
\label{kernel_y2}\\
\frac{G(y)^{-2-6\delta}G(y)^{-2-6\delta}}{Z(y)^{3\delta}Z(z)^{3\delta}xy} 
Q^\prime(x,y,z) &\rightarrow& \frac{G(0)^{-2-6\delta}G(x)^{-2-6\delta}}
{Z(0)^{3\delta}Z(x)^{3\delta}xy} 
\left(\frac{\zeta\cos^2\Theta}{xy}+...\right) \; .
\label{kernel_y3}  
\end{eqnarray}
Furthermore, $z=x+y-2\sqrt{xy}\cos\Theta$ 
has been used and terms proportional to $\cos\Theta$ have been dropped, 
as they either integrate to zero in the continuum or cancel each other in the 
sums on the torus. Each of the expressions 
(\ref{kernel_y1},\ref{kernel_y2},\ref{kernel_y3}) is then the appropriate
term for $j=0$ on the right hand side of the Dyson-Schwinger equations on the 
torus. Clearly one observes, that only a finite ghost mode $G(0)$ avoids 
trouble with divergencies. This is especially true for the kernel $Q^\prime$ of the 
gluon loop, as $Z^{-3\delta}(y\rightarrow 0)$ is more singular than
the simple pole, so this kernel vanishes for small momenta $y$. The other two 
expressions (\ref{kernel_y1}) and (\ref{kernel_y2}) are finite. One is then 
left with the ambiguous quantities $\sin^2\Theta$ and $\cos^2\Theta$ which will 
be replaced by their integrals from zero to $2\pi$ in the calculation at the end 
of this section. The arbitrarines of this procedure is considerably milded by the 
observation that any number plugged in for the trigonometric functions yields the 
same qualitative result at the end of this section. 

Second, we take the limit $z\rightarrow 0$, which on the
torus is identical to $\Theta \rightarrow 0$. The ghost kernel
$\sin^2\Theta/z^2$ alone would certainly diverge as $\Theta \rightarrow 0$, 
but taking into account the power law behaviour $Z(z) \sim z^{2\kappa}$
for the gluon dressing function the integrand is zero in this limit. This is valid for
$\kappa > 0.5$, which is in agreement with the infrared analysis in the continuum.
We therefore may omit the points $z=0$ in the ghost equation.  
The situation is different in the gluon equation where
the kernel of the ghost loop has a finite limit $z\rightarrow 0$: $M(x,x,0)/xy=(\zeta+1)/(2x^2)$. 
Therefore with a finite ghost dressing function $G(0)$ the points $z=0$ in the ghost loop
contribute but no divergencies occur.
In the gluon loop the kernel $Q^\prime$ multiplied by the dressing functions
approaches zero for vanishing momentum $z$ due to the power law behaviour of 
$Z^{-3\delta}(z\rightarrow 0)$. 

To obtain a definite value for $G(0)$ we now investigate the behaviour of
the equations (\ref{ghostbare_app},\ref{gluonbare_app})
in the limit $x\rightarrow0$. The integrands are then given by 
\begin{eqnarray}
\frac{G(y) Z(z)}{xy}K(x,y,z) &\rightarrow& G(y)Z(y)\frac{\sin^2\Theta}{y^2}\; ,\\
\frac{G(y) G(z)}{xy}M(x,y,z) &\rightarrow& G(y)G(y)\left(\frac{1-\zeta\cos^2\Theta}{xy}
+... \right)\; ,
\label{kernel_x}
\end{eqnarray}
where the kernel $Q^\prime$ is of no interest, as we know the gluon loop to be subleading
in the infrared. Clearly, the kernel $M$ of the ghost loop in the gluon 
equation is now singular for $x\rightarrow 0$, 
corresponding to a vanishing gluon dressing function in 
the infrared. This result confirms our working hypothesis that the gluon mode 
$Z(0)$ is not affected by the finite volume of the torus. 
The integrand of the ghost equation is finite up to the point $y=0$. There
the pole in the kernel is canceled by the behavior of the
gluon dressing function $Z(y) \sim y^{2\kappa}$ resulting in a zero for
vanishing momentum $x$ and $y$.

We therefore arrive at a consistent
set of equations for a finite ghost mode $G(0)$ and a vanishing gluon mode $Z(0)$.
\begin{figure}
\centerline{
\epsfig{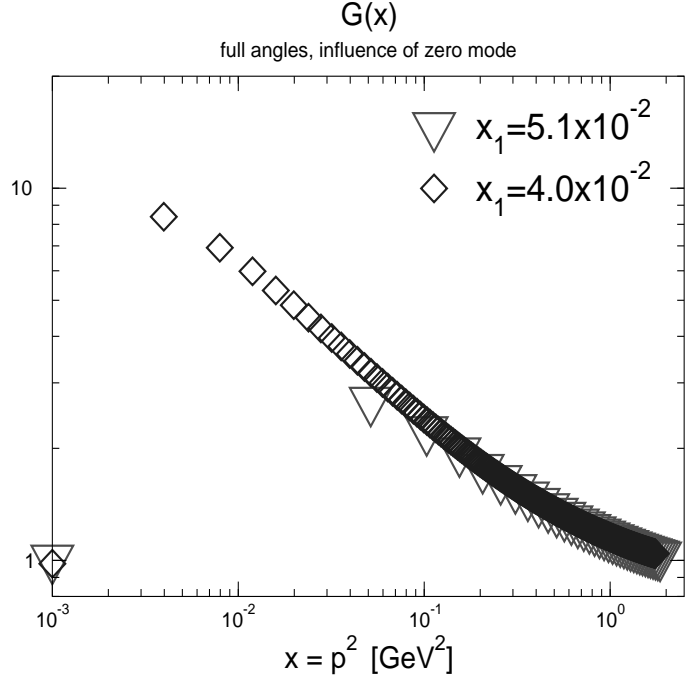}
}
\caption{\label{null.dat}
The ghostmode $G(0)$ compared with the results for finite momentum $x$ on the torus.
For convenience we have kept the logarithmic momentum scale and plotted the zero modes
on the left border of the figure.}
\end{figure}
In Fig.~\ref{null.dat} we show the results for the ghost dressing function
gained on two different volumes on the torus. Within numerical accuracy the values 
of $G(0)$ are the same
for the two volumes. Obviously terms with high loop momentum $y$ 
contribute most to the right hand side of the ghost equation for vanishing momentum $x$.
Furthermore, one observes that the actual value of $G(0)$ is not in accordance with
an extrapolation of the ghost curves to the infrared. There is also no sizeable change of the gluon and the ghost 
dressing function when $G(0)$ is set to zero by hand. This has been done in all 
calculations in the main body of this paper.

Having shown that $G(0)< \infty$ on the torus even in the infinite-volume limit
and assuming that the torus should provide a reasonable infrared
regularization  of physics in flat space-time we conclude that the divergence
of $G(0)$  is very probably due to the torus zero modes of gluons and ghosts.
Noting furthermore that a diverging  $G(0)$ is related to the Zwanziger--Gribov
horizon condition and the Kugo--Ojima confinement criterion this indicates a
direct relation between zero modes, the Gribov horizon and confinement.

\section{Appendix C: Numerical methods for flat Euclidean space-time}

The numerical methods to solve the coupled gluon-ghost system in flat Euclidean
space-time after applying  angular approximations has been described in great
detail in refs.\  \cite{Hauck:1998sm,Atkinson:1998tu}. Note that different
numerical techniques have been employed in refs.\
\cite{vonSmekal:1997is,Hauck:1998sm} and ref.\
\cite{Atkinson:1998tu}. We have used all these techniques to
verify the independence of our solutions from details of the numerical
treatment.

First, we discuss the numerical renormalization process. As we know
the characteristics of the solutions {\it a priori} we may exploit the
possibility of independent subtractions of the gluon and the ghost equation.
The procedure followed in ref.\ \cite{vonSmekal:1997is} was to eliminate
renormalization constants $Z_3$ and $\tilde{Z}_3$ by subtracting the equation
at the renormalization scale $s=\mu^2$. The results presented in sect.~IV.A are
generated by solving the equations
\begin{eqnarray}
\frac 1 {Z(x)} - \frac 1 {Z(s_Z)} &=&  \frac {g^2 N_c}{48\pi^2}
\left(
G(x) \int_0^x \frac{dy}{x} \left( -\frac{y^2}{x^2} + \frac{3y}{2x}
\right) G(y) + \int_x^{\Lambda^2} \frac{dy}{2y} G^2(y) \right. \nonumber\\
&&\left. -G(s_Z) \int_0^{s_Z} \frac{dy}{s_Z} \left( -\frac{y^2}{s_Z^2} + 
\frac{3y}{2s_Z}
\right) G(y) + \int_{s_Z}^{\Lambda^2} \frac{dy}{2y} G^2(y)
\right) ,
\\
\frac 1 {G(x)} - \frac 1 {G(s_G)} &=&  - \frac 9 4 \frac {g^2 N_c}{48\pi^2}
\left(
 Z(x) \int_0^x \frac{dy}{x} \frac{y}{x} G(y)
+ \int_x^{\Lambda^2} \frac{dy}{y} Z(y) G(y) \right. \nonumber\\
&& \left. Z(s_G) \int_0^{s_G} \frac{dy}{s_G} \frac{y}{s_G} G(y)
+ \int_{s_G}^{\Lambda^2} \frac{dy}{y} Z(y) G(y)
\right) ,
\end{eqnarray}
for the truncation scheme of sect.~II.B and analogous equations for the scheme
of sect.~II.A thereby introducing subtraction points $s_Z$ and $s_G$.
The parts of the integrals  from $y=0$ to an infrared matching point 
$\epsilon$ are carried out analytically, with the dressing functions
in the integral beeing replaced by their leading infrared behaviour
(\ref{ZABG})
$$
Z(x) = Ax^{2 \kappa}, \:\:\:\: G(x) = B x^{- \kappa} .
$$
The infrared analysis leads to fixed values  $\kappa$ and  
$\alpha_c = \alpha(0) =(\frac{g^2N_c}{12\pi}) AB^2$, 
so $A$ is left as a free parameter in the infrared expansion. For
numerical reasons it is convenient to choose the subtraction points $s_G=0$ and
$s_Z$ to lie above the maximum of $Z(x)$. 
The two input parameters that determine the renormalization point
where we require the normalization conditions $G(s=\mu^2)=Z(s=\mu^2)=1$ 
are then $A$ and the value the gluon dressing function
at its subtraction point, {\it i.e.} $Z(s_Z)$.

The equations are solved using the Newton iteration method thereby generating
values for the propagator functions. These are then plugged into the
unsubtracted equations to obtain the respective values of the renormalization 
constants $Z_3$ and $\tilde{Z}_3$ for a given cutoff $\Lambda$ and a given
renormalization scale $\mu$ which allows finally to determine  the  numerical
solutions respecting the normalization conditions $G(s=\mu^2)=Z(s=\mu^2)=1$.

\end{appendix}

\end{document}